\documentclass[%
 reprint,
superscriptaddress,
 amsmath,amssymb,
 aps,
prl,
]{revtex4-1}

\usepackage{graphicx}
\usepackage{dcolumn}
\usepackage{bm}
\usepackage{hyperref}
\usepackage{textcomp}
\usepackage[utf8]{inputenc}
\usepackage{xcolor}

\begin{document}

\newcommand{\sd}[1]{\textcolor{red}{#1}}

\newcommand{\NISTTF}{Time and Frequency Division, NIST, Boulder CO 80305 USA}
\newcommand{\CUPhysics}{Department of Physics, University of Colorado, Boulder CO 80305 USA}
\newcommand{\ICFO}{ICFO -- Institut de Ci\`encies Fot\`oniques, The Barcelona Institute of Science and Technology, 08860 Castelldefels (Barcelona), Spain}
\newcommand{\ICREA}{ICREA, Pg. Lluís Companys 23, 08010 Barcelona, Spain}
\newcommand{\BAE}{BAE Systems, Nashua NH USA}
\newcommand{\Campinas}{Instituto de Fisica Gleb Wataghin, Universidade Estadual de Campinas, Campinas, SP, 13083-859, Brazil}
\newcommand{\EqContrib}{These authors contributed equally to this work.}
\newcommand{\micron}{\textmu{}m}
\preprint{APS/123-QED}

\title{Infrared electric-field sampled frequency comb spectroscopy}

\author{Abijith S. Kowligy}
\thanks{\EqContrib}
\affiliation{\NISTTF}
\author{Henry Timmers}%
\thanks{\EqContrib}
\affiliation{\NISTTF}
\author{Alex Lind}
\thanks{\EqContrib}
\affiliation{\NISTTF}
\affiliation{\CUPhysics}
\author{Ugaitz Elu}
\affiliation{\ICFO}
\author{Flavio C. Cruz}
\affiliation{\NISTTF}
\affiliation{\Campinas}
\author{Peter G. Schunemann}
\affiliation{\BAE}
\author{Jens Biegert}
\affiliation{\ICFO}
\affiliation{\ICREA}
\author{Scott A. Diddams}
\affiliation{\NISTTF}
\affiliation{\CUPhysics}

\date{\today}

\begin{abstract}
Molecular spectroscopy in the mid-infrared portion of the electromagnetic spectrum (3\textendash{}25 \micron) has been a cornerstone interdisciplinary analytical technique widely adapted across the biological, chemical, and physical sciences.  Applications range from understanding mesoscale trends in climate science via atmospheric monitoring to microscopic investigations of cellular biological systems via protein characterization. Here, we present a compact and comprehensive approach to infrared spectroscopy incorporating the development of broadband laser frequency combs across 3\textendash{}27 \micron, encompassing the entire mid-infrared, and direct electric-field measurement of the corresponding near single-cycle infrared pulses of light. Utilizing this unified apparatus for high-resolution and accurate frequency comb spectroscopy, we present the infrared spectra of important atmospheric compounds such as ammonia and carbon dioxide in the molecular fingerprint region. To further highlight the ability to study complex biological systems, we present a broadband spectrum of a monoclonal antibody reference material consisting of more than 20,000 atoms. The absorption signature resolves the amide I and II vibrations, providing a means to study secondary structures of proteins. The approach described here, operating at the boundary of ultrafast physics and precision spectroscopy, provides a table-top solution and a widely adaptable technique impacting both applied and fundamental scientific studies. 
\end{abstract}

\maketitle

Over the past fifty years, advances in infrared (IR) technologies have played a vital role in shaping our understanding of the internal structure of molecular compounds via linear and nonlinear spectroscopic methods \cite{stuart_infrared_2004}. Specifically, the advent of the Fourier-transform infrared spectrometer (FTIR) has enabled studying vital chemical and biological processes as nearly all the relevant molecules exhibit unique absorption signatures corresponding to their ro-vibrational degrees of freedom in the molecular fingerprint region \cite{griffiths_fourier_2007}. Probing the properties of matter in this scientifically critical spectral region has been fruitful for a wide variety of scientific and industrial applications including atmospheric monitoring \cite{noziere_molecular_2015}, food quality control \cite{sun_infrared_2009}, and conservation and composition studies of natural and synthetic materials \cite{chukanov_infrared_2016,derrick_infrared_1999}. Importantly, studying biological systems via IR spectroscopy has revealed structural and conformational information in complex organic compounds such as proteins, which is of fundamental significance for applications such as drug synthesis and delivery \cite{barth_infrared_2007}. While the broadband incoherent radiation in conventional FTIR has been sufficient for many of these studies, several fundamental scientific applications and next-generation analytical techniques require high-brightness, coherent sources of MIR light, along with commensurate rapid acquisition and high signal-to-noise ratio (SNR) measurement methodologies \cite{pires_ultrashort_2015}. 

Recently, coherent IR spectro-imaging of biological specimens has been proposed as a complementary technique in clinical pathology that could result in stain-free histology and label-free cellular analysis \cite{katon_infrared_1996,fernandez_infrared_2005,bellisola_infrared_2011,clemens_vibrational_2014}.  Moreover, near-field analytical techniques such as atomic force microscopy in the infrared (AFM-IR) enable nanoscale imaging of single proteins \cite{keilmann_near-field_2004,dazzi_afm-ir:_2017}. In addition to biologically relevant molecular compounds, near-field probing of synthetic polymers and exotic quantum matter using techniques such as scattering scanning near-field optical microscopy (sSNOM) also reveal fundamental structural information useful for material science \cite{bechtel_ultrabroadband_2014}. These revolutionary applications place exacting criteria on the combination of the infrared sources and detection apparatus\textemdash{}requiring high-brightness coherent broadband light and high-SNR readout of the IR spectra. This has meant accessing IR synchrotron beamlines and using cryogenically-cooled detectors for low-noise operation, thereby placing significant constraints on accessibility, mobility, and cost \cite{khatib_far_2018}. 

\begin{figure}[t!]
\centering
\includegraphics[width=\linewidth]{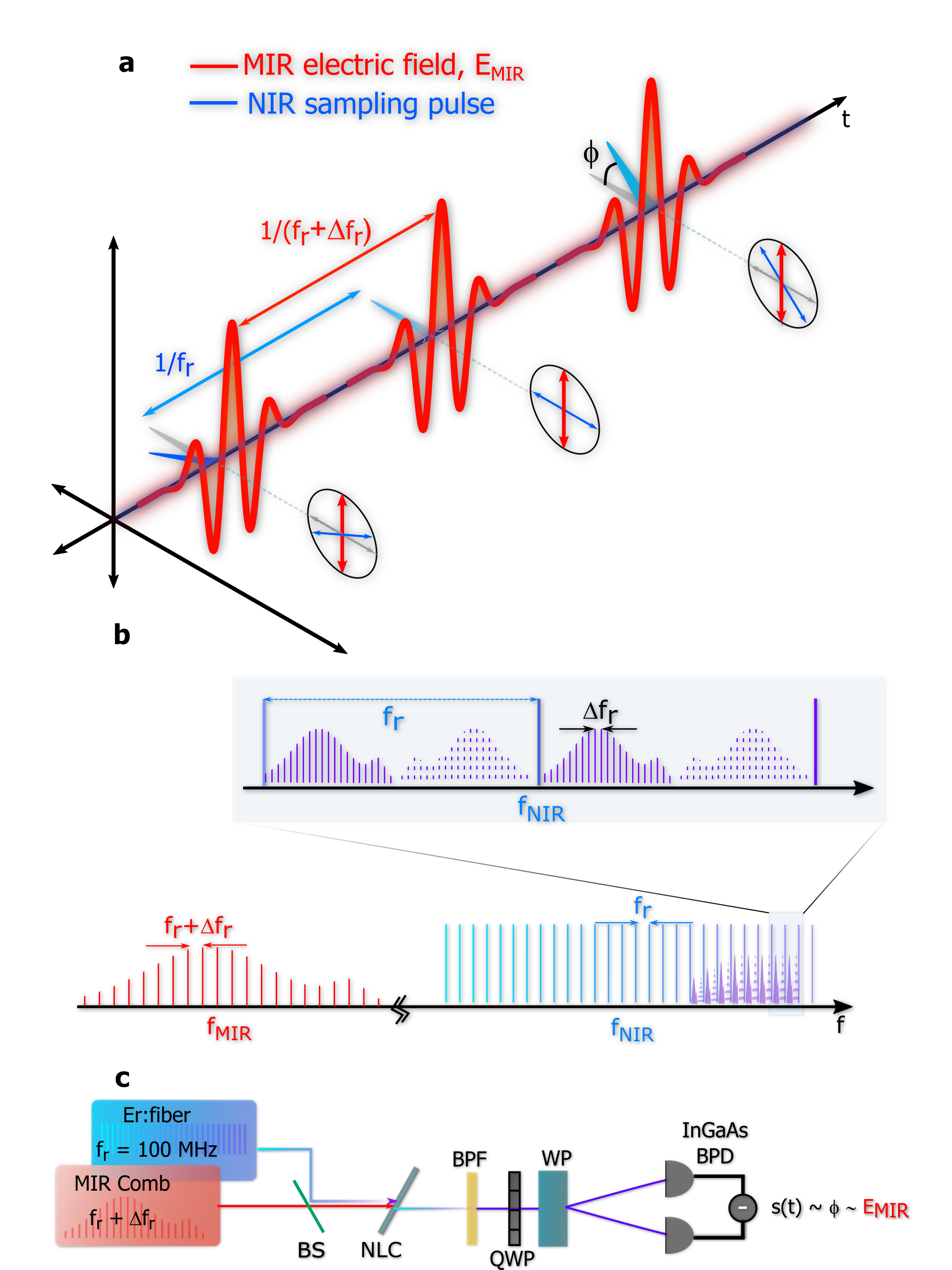}
\caption{Direct nonlinear optical readout of MIR electric fields. (a) A MIR electric field (red) induces a nonlinear polarization rotation on a NIR sampling pulse (blue), whose temporal duration is small compared to the MIR optical cycle. The polarization rotation,$\phi$, occurs via sum-frequency generation (SFG) in an electro-optic crystal and is proportional to the MIR electric amplitude, $E_\text{MIR}$. Due to a small offset in the repetition rates of the NIR ($f_r$) and MIR ($f_r + \Delta f_r$) pulse trains, the measurement is rendered periodic at a rate $\Delta f_{r}$. (b) In the frequency domain, sum-frequency components are folded into the free-spectral range (FSR) of the high-frequency region of the sampling pulse. In this spectral region, owing to the offset repetition rates, the entire MIR spectrum lies within every Nyquist zone ($f_r/2$). (c) The NIR and MIR pulses are combined at a Ge beam-spliitter (BS), and owing to birefringence phase-matching in the electro-optic crystal (NLC), the sum-frequency components are orthogonally polarized to the sampling pulse. Spectral filtering via a band-pass filter (BPF) and subsequent polarization-resolving ellipsometry using a quarter wave-plate (QWP), Wollaston prism (WP), and balanced InGaAs photodetectors yields a signal, $s(t)$, that is proportional to the MIR electric field. }
\label{fig:concept}
\end{figure}

Circumventing these limitations, we present an ultrabroadband source of laser frequency combs spanning 3--27 \textmu{}m, which produces near-single-cycle pulses of light. Importantly, due to the enhanced peak power in such pulses, we use a nonlinear optical readout based on electro-optic sampling (EOS) to directly measure the IR electric fields with room-temperature telecom-grade InGaAs photodiodes. This high-SNR electric field measurement enables high-sensitivity and broadband IR spectroscopy with a large dynamic range. Moreover, we incorporate a dual frequency comb implementation of EOS (henceforth dual-comb EOS) \cite{bartels_ultrafast_2007}, enabling electric-field measurement at a rate of 50 Hz, with high temporal (5 fs) and spectral (100 MHz) resolutions. This unique combination of video-rate acquisition and high resolution represents a significant improvement (by at least an order-of-magnitude) over prior MIR-EOS demonstrations  \cite{sell_field-resolved_2008,riek_direct_2015,pupeza_field-resolved_2017}. This cohesive approach is a compact, table-top alternative to large IR synchrotron beamlines and broadband incoherent thermal sources in conventional FTIR systems. 



Generation of the broadband infrared light relies on nonlinear frequency conversion processes driven by stabilized erbium fiber frequency combs. The frequency comb architecture enables transferring the exceptional frequency accuracy and stability of the 100~MHz~spaced NIR comb teeth to all portions of the electromagnetic spectrum \cite{schliesser_mid-infrared_2012}. Harnessing these advantages, dual frequency comb spectroscopy (DCS) has been demonstrated as a high-resolution, fast-acquisition alternative to FTIR \cite{coddington_dual-comb_2016}. Recently, high-resolution quantitative spectroscopy has been demonstrated in both the 3\textendash{}5~\textmu{}m and the 6\textendash12~\textmu{}m portions of the mid-infrared using DCS \cite{kara_dual-comb_2017,maidment_long-wave_2018,muraviev_massively_2018,ycas_high-coherence_2018,timmers_molecular_2018}. However, for low-noise operation, DCS experiments have all required cryogenically cooled infrared photodetectors, which exhibit a limited range of linearity\textemdash{}typically limited to sub-milliwatt optical powers. This fundamentally limits the achievable signal-to-noise ratio \cite{newbury_sensitivity_2010}. 

\begin{figure*}[t]
\centering
\includegraphics[width=\linewidth]{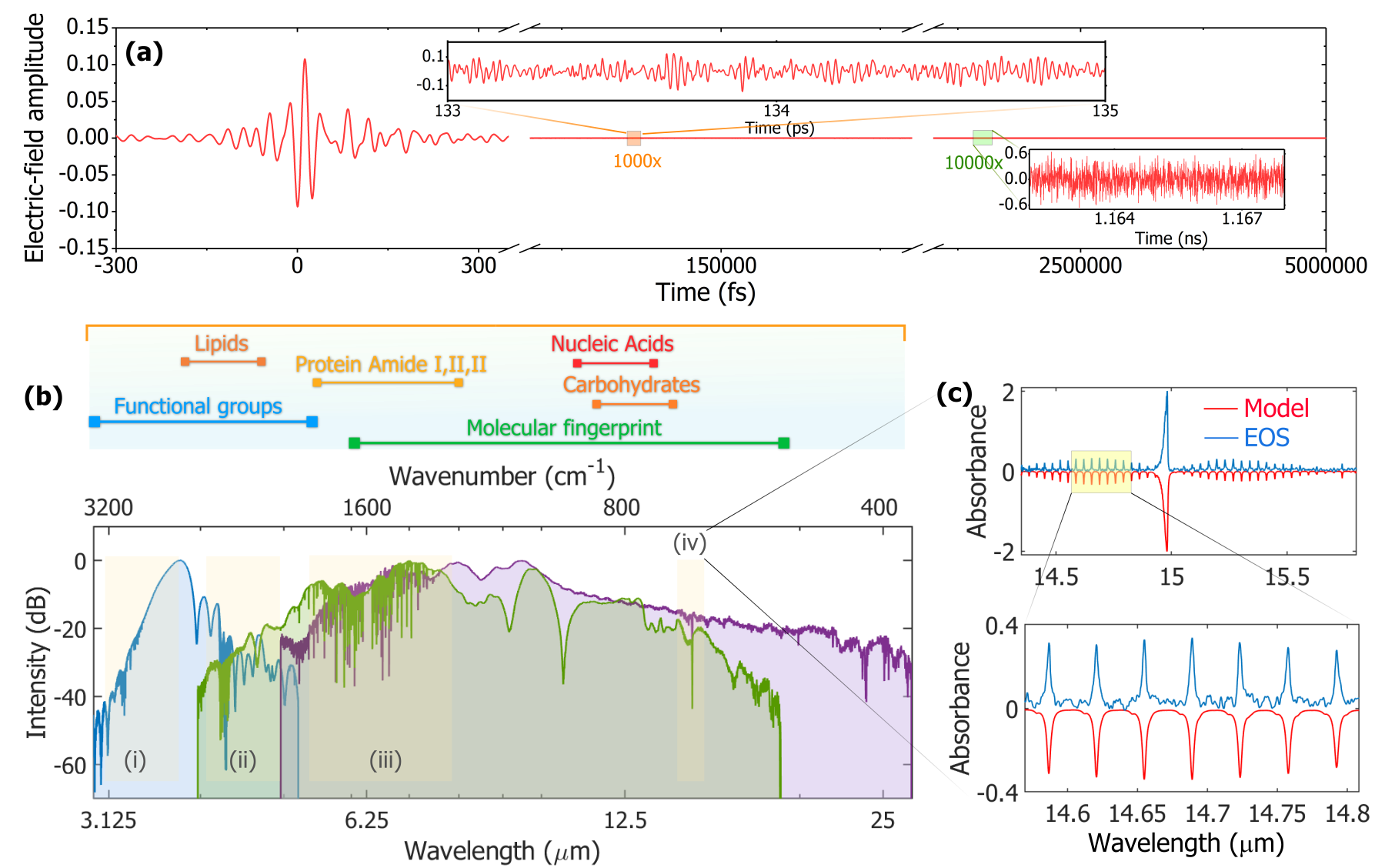}
\caption{Electro-optically sampled ultrabroadband infrared spectra. (a) An electro-optically sampled 1.2-cycle MIR electric field, oscillating at 39~THz, is shown with 5 fs resolution extending over a 10 ns time window. The dual frequency comb measurement enables observation of the trailing molecular free-induction decay, 135 ps away from the centerburst (inset, top). High SNR is observed with the noise being suppressed by a factor of $\sim 10^{4}$ (inset, bottom) (b) Ultrabroadband MIR spectra corresponding to the Fourier transform of the measured electric fields from PPLN (blue), OP-GaP (green), and GaSe ( purple).  Absorption signatures of (i) C-H stretch vibrations for CH$_4$ and C$_2$H$_6$ (ii) anti-symmetric  C-O  stretch  for  atmospheric  CO$_2$ (iii)  atmospheric  H$_2$O  and  (iv)  the  bending  vibration  of atmospheric CO$_2$ are all seen. (c)The CO$_2$ bending vibration in the 14\textendash{}16~\micron~region, averaged for 22 minutes, is presented with 1~GHz resolution (blue) and quantitatively compared to a theoretical model from the HITRAN2012 database (red). }
\label{fig:spectra}
\end{figure*}

Here, we utilize the combination of difference-frequency generation (DFG) and sum-frequency generation (SFG) to generate and detect the infrared frequency combs. The generation of the infrared light is based on intra-pulse DFG in quadratic nonlinear media, which provides ultrashort infrared pulses with an inherently stabilized carrier-envelope phase (CEP) \cite{baltuska_controlling_2002}. The CEP stability enables a direct nonlinear optical readout of the MIR electric field via electro-optic sampling \cite{wu_freespace_1995,sell_field-resolved_2008}. In EOS, the MIR electric field, with a carrier frequency $f_{c}$, modulates the polarization of an ultra-short NIR sampling pulse in an electro-optic crystal, as shown in Fig.~\ref{fig:concept}(a). The polarization modulation is directly proportional to the MIR field amplitude and occurs due to SFG between the NIR and MIR pulses. The sum-frequency NIR components and the input sampling pulse are orthogonally polarized, which provides a polarization rotation \cite{sell_field-resolved_2008,pupeza_high-power_2015}. As the time delay is scanned between the two pulses, a complete measurement of the electric field is made. 



Previous EOS experiments have shown its advantages for spectroscopy \cite{sell_field-resolved_2008} and measurements with spectral dynamic range $>$110 dB have been reported \cite{pupeza_field-resolved_2017}. Additionally, it has improved sensitivity relative to FTS \cite{KrauszConf2} with the ability to directly measure vacuum fluctuations\cite{riek_direct_2015}.  However, in all such cases the delay between the NIR and MIR pulses is varied using a mechanical translation stage, restricting the range of measurement and acquisition time to centimeter- and second-scales, respectively. 
On the other hand, in dual-comb EOS, fast acquisition of high-resolution infrared spectra is feasible as the sampling pulse and the MIR light are acquired from two different, mutually phase-locked Er:fiber laser frequency combs with a small offset in repetition rates ($\Delta f_r$ = 50 Hz) as shown in Fig.~\ref{fig:concept}(a). Unlike conventional frequency-comb-enabled linear optical sampling \cite{coddington_rapid_2009,coddington_coherent_2009,coddington_time-domain_2010}, the down-sampled comb folds the entire MIR spectrum into each FSR of the NIR comb in dual frequency comb EOS (Fig.~\ref{fig:concept}(b),\cite{stead_method_2012}). This is similar to the case of the stimulated Raman experiments performed with a dual comb configuration \cite{ideguchi_coherent_2013}. Thus, the coherent multi-heterodyne beating occurs in the NIR, with the sampling pulse serving as a local-oscillator comb. For this reason, the sampling pulse must have a bandwidth greater than the highest MIR frequency component to be sampled \cite{riek_direct_2015}. In the time-domain, this means the sampling pulse duration is short compared to the optical cycle $\tau_{\textnormal{MIR}} = 1/f_{c}$. The experimental implementation of EOS is shown in Fig.~\ref{fig:concept}(c). We utilize a 10~fs, 1.55-\textmu{}m sampling pulse \cite{timmers_molecular_2018}, ensuring Nyquist-limited sampling of electric fields corresponding to 3 \textmu{}m (3300 cm$^{-1}$) light. 



\begin{figure*}[t]
\centering
\includegraphics[width=\linewidth]{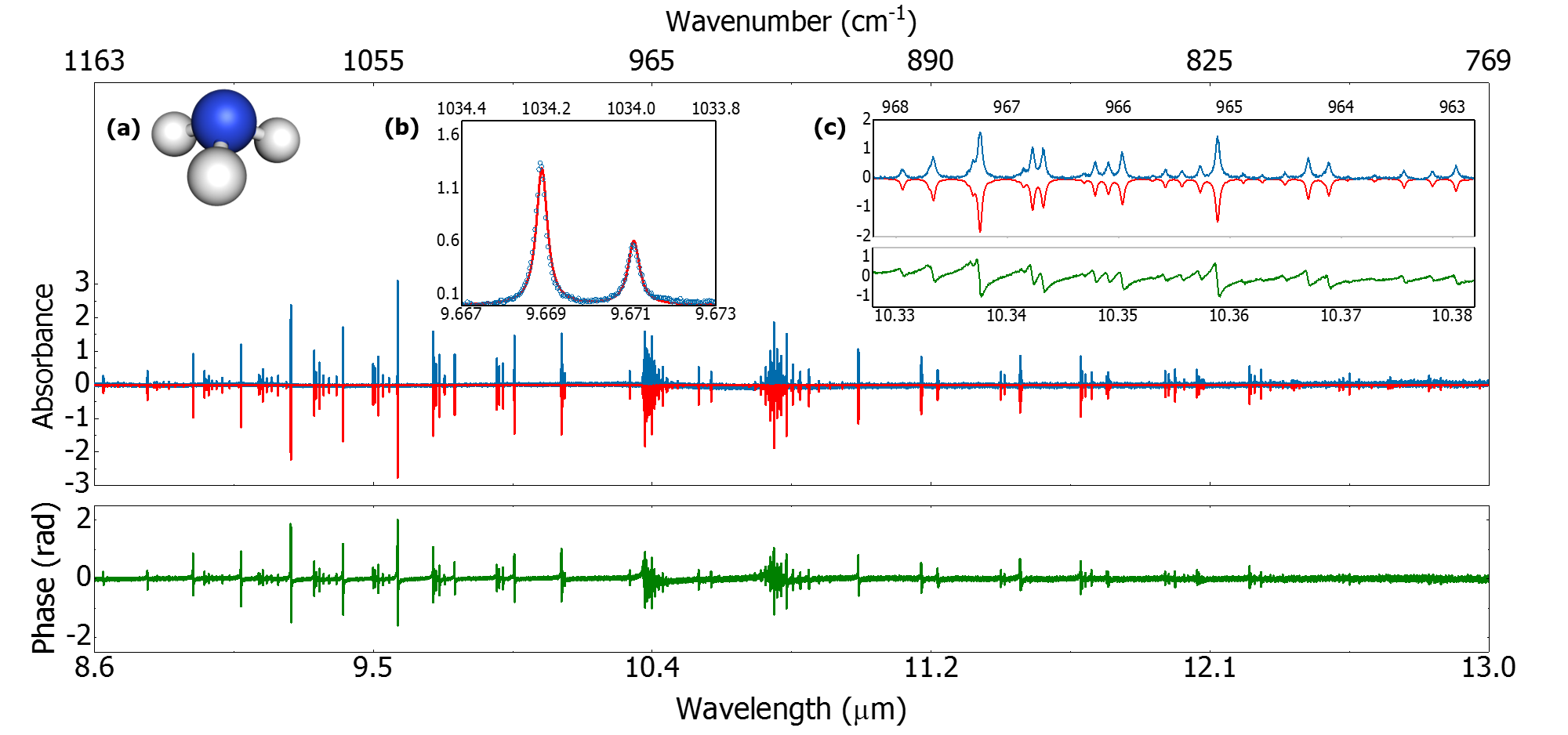}
\caption{High-resolution infrared spectroscopy of gas-phase ammonia (NH$_3$). (a) High-resolution spectroscopy of the $\nu_2$ vibration (A$_1$ symmetric bend) of gas-phase ammonia is measured in the span 8.6\textendash{}13 \micron~region with 100~MHz~resolution via electro-optic sampling of the OP-GaP MIR comb. The averaging time was 88 minutes. (b) The comb-mode resolved spectrum is shown with each comb tooth (blue circles) stabilized to a 10~kHz linewidth. The comparison with the HITRAN2012 (red) is presented. (c) Despite having 30 dB spectral intensity variation in this spectral region, the measured NH$_3$ data compares well to the HITRAN2012 theoretical model due to the high SNR per mode (SNR $\sim$ 65). }
\label{fig:nh3}
\end{figure*}

Coherent infrared radiation is generated via intra-pulse DFG in three different nonlinear crystals\textemdash{}periodically poled lithium niobate (PPLN), orientation-patterned gallium phosphide (OP-GaP), and gallium selenide (GaSe). We leverage quasi phase-matching in PPLN and OP-GaP and birefringence phase-matching in GaSe. Due to favorable dispersion and millimeter-scale crystal propagation lengths, the broadband phase-matching bandwidths are realized in all configurations. Correspondingly, in the time-domain, ultrashort femtosecond MIR pulses are generated. Owing to the enhanced peak power, the nonlinear optical readout of the electric field occurs with signal-to-noise ratio (SNR) exceeding 10$^{4}$ in 20 minutes of averaging, corresponding to a single-shot SNR $\sim 40$.  For the MIR comb generated in OP-GaP, the anomalous dispersion from crystal propagation is compensated by the normal dispersion from the germanium beam-splitter, and results in a near-single-cycle oscillation (Fig.~\ref{fig:spectra}a). The observed time-domain pulse is 1.2 cycles in duration, corresponding to 29 fs, with a carrier wavelength of $\sim$ 7.6 \textmu{}m (1316 cm$^{-1}$).  Unique to dual-comb EOS, the electric field is measured with a 5-fs resolution over a 10-ns window, corresponding to a dynamic range in temporal duration of $>$~60 dB (Fig.~\ref{fig:spectra}a).%

The measurements described here are exceptionally broad in bandwidth, with the spectral coverage rivaling a thermal source in a commercial Fourier-transform infrared spectrometer. The frequency domain SNR scales with power, bandwidth, and spectral resolution and for 100~MHz resolution, we report SNR = 0.9 Hz$^{1/2}$ over a recovered bandwidth of nearly 75 THz. The noise in the system stems from the shot noise of the sampling pulse and excess electrical noise from the balanced photo-detector. High-linearity InGaAs photodetectors would alleviate this issue. Due to the expansive bandwidth in the system, the SNR is smaller compared to dual-comb spectrometers \cite{ycas_high-coherence_2018,timmers_molecular_2018}. However, the figure-of-merit (FOM, $M \times \text{SNR}$), accounting for the number of comb-teeth (M) is comparable to other demonstrations. For the OP-GaP IR comb, corresponding approximately to 10$^6$ comb teeth, the FOM is 10$^6$, which is within a factor of 2 from the best MIR DCS performance. 

Owing to the high SNR in the temporal measurement, the molecular free-induction decay of trace absorbents such as methane and ethane, along with ambient water vapor and carbon dioxide are also obtained via direct electric-field sampled spectroscopy (Fig.~\ref{fig:spectra}b). In particular, 1~GHz (0.033 cm$^{-1}$) resolution vibrational spectra of the P, Q, and R branches of the O\textendash{}C\textendash{}O bending vibration around 15 \textmu{}m (670 cm$^{-1}$) are presented in Fig.~\ref{fig:spectra}c, along with a comparison to theory (HITRAN2012). Notably, this spectral region is beyond the detection range of current state-of-the-art high-speed mercury cadmium telluride (HgCdTe) photodetectors. While 1~GHz resolution is sufficient for pressure-broadened gases, higher resolution is desired for many precision measurements. 

To demonstrate this, we present high-resolution absortion spectra of the $\nu_2$ vibration  (A$_1$ symmetric bend) in ammonia with 100~MHz resolution  (Fig.~\ref{fig:nh3}). Ammonia is a harmful substance widely used in the agricultural and bio-pharmaceutical industries 
\cite{timmer_ammonia_2005,Schliesser:05}, and is an important contributor to atmospheric aerosol formation \cite{kirkby_role_2011}. We observe the splitting in the \textit{Q}-branch due to pyramidal inversion of the NH$_3$ molecule, where the nitrogen atom undergoes room-temperature quantum tunneling through the potential barrier formed by the H$_3$ plane. To acquire this broadband spectrum, the OP-GaP MIR comb is transmitted through a 15-cm-long gas cell, filled with NH$_3$ (7.5 mbar partial pressure) and ambient air (150 mbar background pressure). The transmitted light is detected via EOS. Fourier transformation of the sampled electric field and subtraction of a reference spectrum collected without the gas cell yields the high-resolution absorption spectrum and phase response shown in Fig.~\ref{fig:nh3}a. In this spectrum spanning 8.6\textendash{}13~\micron, 118000 comb teeth are measured. In Fig.~\ref{fig:nh3}b, every absorption feature is resolved with individual comb teeth, which are stabilized to the 10~kHz~level. The frequency accuracy of the stabilized comb teeth is $<10^{-11}$, limited only by the microwave reference for the repetition rate, providing excellent agreement with the HITRAN2012 theoretical model (Fig.~\ref{fig:nh3}c).  

\begin{figure}
\centering
\includegraphics[width=\linewidth]{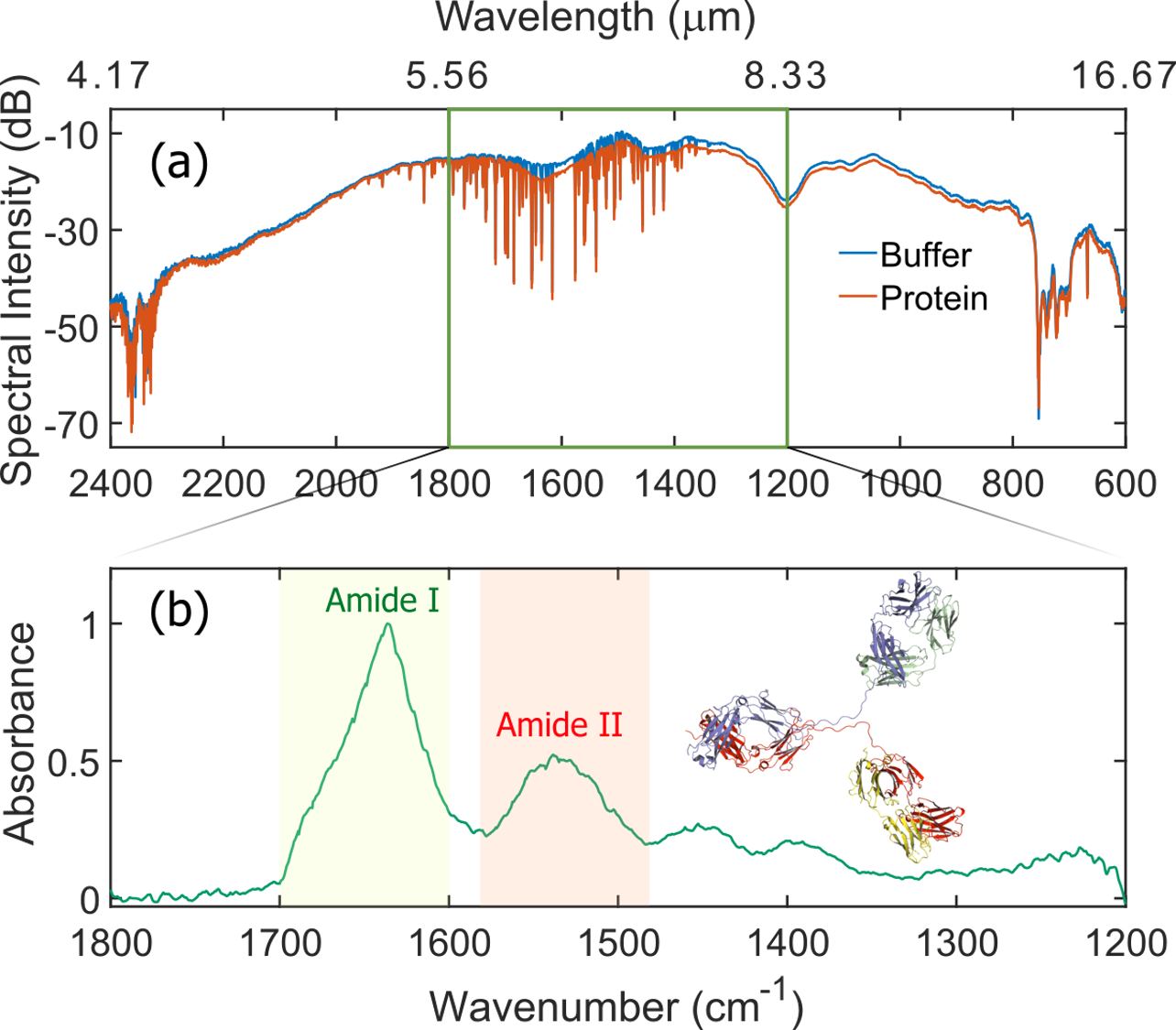}
\caption{NISTmAb infrared spectrum. The 30-GHz resolution (1 cm$^{-1}$) infrared spectrum of the NISTmAb\textemdash{}a monoclonal antibody reference material is obtained in transmission across 600 cm$^{-1}$ (18 THz). The averaging time was 66 minutes. The sample is in the form of a dried film including an L-Histidine buffer on a ZnSe window. The Amide I and II bands are present. The absorption signature is extracted from an ambient background that includes ambient atmospheric water vapor.}
\label{fig:protein}
\end{figure}

Gas-phase frequency comb spectroscopy of small molecules has been a useful diagnostic tool for atmospheric monitoring, probing complex molecules at cryogenic temperatures, and for studying temporal dynamics of chemical reactions \cite{cossel_gas-phase_2017,spaun_continuous_2016,fleisher_mid-infrared_2014}. In contrast, biochemically relevant compounds, such as proteins, often have tens of thousands of atoms, and exhibit interesting infrared spectroscopic signatures in liquid- and solid phases. Here, we use the high-brightness frequency comb to perform infrared spectroscopy of the NISTmAb (Fig.~\ref{fig:protein}), an important reference antibody used in the biopharmaceutical industries \cite{dong_nistmab_2018}. Studying the amide vibrations can aid in determining the protein folding, unfolding, and aggregating mechanisms, important for example, in characterizing drug synthesis and delivery \cite{barth_infrared_2007,baker_ftir-based_2008,baker_using_2014}. We present the absorption spectra of the amide I and II bands in a NISTmAb sample (100 mg/mL), dissolved in L-Histidine buffer (pH: 6.0). The antibody was deposited as a thin film on a 3-mm-thick ZnSe window.  A reference was acquired by depositing the buffer onto a separate ZnSe window.  Due to the broad features, the electric field signal was temporally apodized to provide 1 cm$^{-1}$ (30~GHz) resolution for the 600 cm$^{-1}$-span spectrum. The experiment was conducted in room-temperature ambient conditions and resulting water vapor lines are removed via referencing (Fig.~\ref{fig:protein}, top panel). The center frequency of the amide I band at 1636 cm$^{-1}$, indicative of a $\beta$-sheet structure, and the amide II band at 1549 cm$^{-1}$ agree well with prior measurements and theoretical predictions \cite{gokarn_biophysical_2015}. The IR spectrum in Fig.~\ref{fig:protein} is the average of three measurements, corresponding to a total averaging time of $\sim$ 66 minutes. Notably, the NISTmAb consists of greater than 20,000 atoms, making it one of the largest complex molecules probed with frequency comb spectroscopy.


In conclusion, we presented a high-brightness ultrafast source of mid-infrared radiation and presented a versatile detection technique using dual frequency comb electro-optic sampling. Inheriting the robustness and simplicity from the near-infrared pulses derived from erbium fiber lasers, the compact source provides a table-top footprint for IR spectroscopy in the molecular fingerprint region and beyond. Furthermore, the use of room-temperature InGaAs photodetectors renders this system flexible and reliable for field-based IR spectroscopy and metrology. Along with precision, high-resolution spectroscopy of atmospheric carbon dioxide and ammonia gas in a spectral region beyond the reach of high-speed HgCdTe detectors, we presented the IR absorption spectrum of the NISTmAb with resolved amide vibrations, crucial for understanding and characterizing its secondary structure. As next steps, we anticipate that combined with novel imaging techniques such as AFM-IR and sSNOM, the apparatus described here can enable in-situ spectro-imaging of biological specimens. Moreover, the near single-cycle infrared pulses can serve as robust seeds for optical parametric chirped-pulse amplifiers in strong-field physics \cite{wolter_strong-field_2015}. 

We thank John Schiel and Trina Mouchanoir for providing the NISTmAb samples and Daniel Lesko for providing the NH$_{3}$ sample. We thank David Carlson, Nathan Newbury, Marissa Weichman, Curtis Meuse, and Kevin Cossel for providing useful comments on the manuscript, and Albrecht Bartels and Alfred Leitenstorfer for advice on the experiments. The funding for this work was provided by Defense Advanced Research Projects Agency (DARPA), National Institute of Standards and Technology (NIST), National Science Foundation (NSF) (1708743), National Research Council (NRC), and the Air Force Office of Scientific Research (AFOSR) (FA9550-16-1-0016). U.E. and J.B. acknowledge funding through the ``Severo Ochoa” Programme for Centres of Excellence in R\&D (SEV-2015-0522), MINECO ``FIS2017-89536-P", AGAUR ``2017 SGR 1639" and ERC Advanced Grant ``TRANSFORMER", (788218).This work is a contribution of the United States government and is not subject to copyright in the United States of America.

\bibliography{Zotero}

\begin{thebibliography}{48}%
\makeatletter
\providecommand \@ifxundefined [1]{%
 \@ifx{#1\undefined}
}%
\providecommand \@ifnum [1]{%
 \ifnum #1\expandafter \@firstoftwo
 \else \expandafter \@secondoftwo
 \fi
}%
\providecommand \@ifx [1]{%
 \ifx #1\expandafter \@firstoftwo
 \else \expandafter \@secondoftwo
 \fi
}%
\providecommand \natexlab [1]{#1}%
\providecommand \enquote  [1]{``#1''}%
\providecommand \bibnamefont  [1]{#1}%
\providecommand \bibfnamefont [1]{#1}%
\providecommand \citenamefont [1]{#1}%
\providecommand \href@noop [0]{\@secondoftwo}%
\providecommand \href [0]{\begingroup \@sanitize@url \@href}%
\providecommand \@href[1]{\@@startlink{#1}\@@href}%
\providecommand \@@href[1]{\endgroup#1\@@endlink}%
\providecommand \@sanitize@url [0]{\catcode `\\12\catcode `\$12\catcode
  `\&12\catcode `\#12\catcode `\^12\catcode `\_12\catcode `\%12\relax}%
\providecommand \@@startlink[1]{}%
\providecommand \@@endlink[0]{}%
\providecommand \url  [0]{\begingroup\@sanitize@url \@url }%
\providecommand \@url [1]{\endgroup\@href {#1}{\urlprefix }}%
\providecommand \urlprefix  [0]{URL }%
\providecommand \Eprint [0]{\href }%
\providecommand \doibase [0]{http://dx.doi.org/}%
\providecommand \selectlanguage [0]{\@gobble}%
\providecommand \bibinfo  [0]{\@secondoftwo}%
\providecommand \bibfield  [0]{\@secondoftwo}%
\providecommand \translation [1]{[#1]}%
\providecommand \BibitemOpen [0]{}%
\providecommand \bibitemStop [0]{}%
\providecommand \bibitemNoStop [0]{.\EOS\space}%
\providecommand \EOS [0]{\spacefactor3000\relax}%
\providecommand \BibitemShut  [1]{\csname bibitem#1\endcsname}%
\let\auto@bib@innerbib\@empty
\bibitem [{\citenamefont {Stuart}(2004)}]{stuart_infrared_2004}%
  \BibitemOpen
  \bibfield  {author} {\bibinfo {author} {\bibfnamefont {B.}~\bibnamefont
  {Stuart}},\ }\href@noop {} {\emph {\bibinfo {title} {Infrared {Spectroscopy}:
  {Fundamentals} and {Applications}}}},\ \bibinfo {edition} {1st}\ ed.,\
  Analytical {Techniques} in the {Sciences}\ (\bibinfo  {publisher} {Wiley
  Interscience},\ \bibinfo {year} {2004})\BibitemShut {NoStop}%
\bibitem [{\citenamefont {Griffiths}\ and\ \citenamefont
  {de~Haseth}(2007)}]{griffiths_fourier_2007}%
  \BibitemOpen
  \bibfield  {author} {\bibinfo {author} {\bibfnamefont {P.~G.}\ \bibnamefont
  {Griffiths}}\ and\ \bibinfo {author} {\bibfnamefont {J.~A.}\ \bibnamefont
  {de~Haseth}},\ }\href@noop {} {\emph {\bibinfo {title} {Fourier transform
  infrared spectrometry}}},\ \bibinfo {edition} {2nd}\ ed.\ (\bibinfo
  {publisher} {Wiley Interscience},\ \bibinfo {year} {2007})\BibitemShut
  {NoStop}%
\bibitem [{\citenamefont {Nozi{\`e}re}\ \emph {et~al.}(2015)\citenamefont
  {Nozi{\`e}re}, \citenamefont {Kalberer}, \citenamefont {Claeys},
  \citenamefont {Allan}, \citenamefont {D{\textquoteright}Anna}, \citenamefont
  {Decesari}, \citenamefont {Finessi}, \citenamefont {Glasius}, \citenamefont
  {Grgi{\'c}}, \citenamefont {Hamilton}, \citenamefont {Hoffmann},
  \citenamefont {Iinuma}, \citenamefont {Jaoui}, \citenamefont {Kahnt},
  \citenamefont {Kampf}, \citenamefont {Kourtchev}, \citenamefont {Maenhaut},
  \citenamefont {Marsden}, \citenamefont {Saarikoski}, \citenamefont
  {Schnelle-Kreis}, \citenamefont {Surratt}, \citenamefont {Szidat},
  \citenamefont {Szmigielski},\ and\ \citenamefont
  {Wisthaler}}]{noziere_molecular_2015}%
  \BibitemOpen
  \bibfield  {author} {\bibinfo {author} {\bibfnamefont {B.}~\bibnamefont
  {Nozi{\`e}re}}, \bibinfo {author} {\bibfnamefont {M.}~\bibnamefont
  {Kalberer}}, \bibinfo {author} {\bibfnamefont {M.}~\bibnamefont {Claeys}},
  \bibinfo {author} {\bibfnamefont {J.}~\bibnamefont {Allan}}, \bibinfo
  {author} {\bibfnamefont {B.}~\bibnamefont {D{\textquoteright}Anna}}, \bibinfo
  {author} {\bibfnamefont {S.}~\bibnamefont {Decesari}}, \bibinfo {author}
  {\bibfnamefont {E.}~\bibnamefont {Finessi}}, \bibinfo {author} {\bibfnamefont
  {M.}~\bibnamefont {Glasius}}, \bibinfo {author} {\bibfnamefont
  {I.}~\bibnamefont {Grgi{\'c}}}, \bibinfo {author} {\bibfnamefont {J.~F.}\
  \bibnamefont {Hamilton}}, \bibinfo {author} {\bibfnamefont {T.}~\bibnamefont
  {Hoffmann}}, \bibinfo {author} {\bibfnamefont {Y.}~\bibnamefont {Iinuma}},
  \bibinfo {author} {\bibfnamefont {M.}~\bibnamefont {Jaoui}}, \bibinfo
  {author} {\bibfnamefont {A.}~\bibnamefont {Kahnt}}, \bibinfo {author}
  {\bibfnamefont {C.~J.}\ \bibnamefont {Kampf}}, \bibinfo {author}
  {\bibfnamefont {I.}~\bibnamefont {Kourtchev}}, \bibinfo {author}
  {\bibfnamefont {W.}~\bibnamefont {Maenhaut}}, \bibinfo {author}
  {\bibfnamefont {N.}~\bibnamefont {Marsden}}, \bibinfo {author} {\bibfnamefont
  {S.}~\bibnamefont {Saarikoski}}, \bibinfo {author} {\bibfnamefont
  {J.}~\bibnamefont {Schnelle-Kreis}}, \bibinfo {author} {\bibfnamefont
  {J.~D.}\ \bibnamefont {Surratt}}, \bibinfo {author} {\bibfnamefont
  {S.}~\bibnamefont {Szidat}}, \bibinfo {author} {\bibfnamefont
  {R.}~\bibnamefont {Szmigielski}}, \ and\ \bibinfo {author} {\bibfnamefont
  {A.}~\bibnamefont {Wisthaler}},\ }\href {\doibase 10.1021/cr5003485}
  {\bibfield  {journal} {\bibinfo  {journal} {Chemical Reviews}\ }\textbf
  {\bibinfo {volume} {115}},\ \bibinfo {pages} {3919} (\bibinfo {year}
  {2015})}\BibitemShut {NoStop}%
\bibitem [{\citenamefont {Sun}(2009)}]{sun_infrared_2009}%
  \BibitemOpen
  \bibfield  {author} {\bibinfo {author} {\bibfnamefont {D.-W.}\ \bibnamefont
  {Sun}},\ }\href@noop {} {\emph {\bibinfo {title} {Infrared {Spectroscopy} for
  {Food} {Quality} {Analysis} and {Control}}}},\ \bibinfo {edition} {1st}\ ed.\
  (\bibinfo  {publisher} {Elsevier},\ \bibinfo {year} {2009})\BibitemShut
  {NoStop}%
\bibitem [{\citenamefont {Chukanov}\ and\ \citenamefont
  {Chervonnyi}(2016)}]{chukanov_infrared_2016}%
  \BibitemOpen
  \bibfield  {author} {\bibinfo {author} {\bibfnamefont {N.~V.}\ \bibnamefont
  {Chukanov}}\ and\ \bibinfo {author} {\bibfnamefont {A.~D.}\ \bibnamefont
  {Chervonnyi}},\ }\href@noop {} {\emph {\bibinfo {title} {Infrared
  {Spectroscopy} of {Minerals} and {Related} {Compounds}}}},\ \bibinfo
  {edition} {1st}\ ed.,\ Springer {Minerology}\ (\bibinfo  {publisher}
  {Springer International Publishing},\ \bibinfo {year} {2016})\BibitemShut
  {NoStop}%
\bibitem [{\citenamefont {Derrick}\ \emph {et~al.}(1999)\citenamefont
  {Derrick}, \citenamefont {Stulik},\ and\ \citenamefont
  {Landry}}]{derrick_infrared_1999}%
  \BibitemOpen
  \bibfield  {author} {\bibinfo {author} {\bibfnamefont {M.}~\bibnamefont
  {Derrick}}, \bibinfo {author} {\bibfnamefont {D.}~\bibnamefont {Stulik}}, \
  and\ \bibinfo {author} {\bibfnamefont {J.~M.}\ \bibnamefont {Landry}},\
  }\href@noop {} {\emph {\bibinfo {title} {Infrared {Spectroscopy} in
  {Conservation} {Science}}}},\ Scientific {Tools} for {Conservation}\
  (\bibinfo  {publisher} {The Getty Conservation Institute},\ \bibinfo {year}
  {1999})\BibitemShut {NoStop}%
\bibitem [{\citenamefont {Barth}(2007)}]{barth_infrared_2007}%
  \BibitemOpen
  \bibfield  {author} {\bibinfo {author} {\bibfnamefont {A.}~\bibnamefont
  {Barth}},\ }\href {\doibase 10.1016/j.bbabio.2007.06.004} {\bibfield
  {journal} {\bibinfo  {journal} {Biochimica et Biophysica Acta (BBA) -
  Bioenergetics}\ }\textbf {\bibinfo {volume} {1767}},\ \bibinfo {pages} {1073}
  (\bibinfo {year} {2007})}\BibitemShut {NoStop}%
\bibitem [{\citenamefont {Pires}\ \emph {et~al.}(2015)\citenamefont {Pires},
  \citenamefont {Baudisch}, \citenamefont {Sanchez}, \citenamefont {Hemmer},\
  and\ \citenamefont {Biegert}}]{pires_ultrashort_2015}%
  \BibitemOpen
  \bibfield  {author} {\bibinfo {author} {\bibfnamefont {H.}~\bibnamefont
  {Pires}}, \bibinfo {author} {\bibfnamefont {M.}~\bibnamefont {Baudisch}},
  \bibinfo {author} {\bibfnamefont {D.}~\bibnamefont {Sanchez}}, \bibinfo
  {author} {\bibfnamefont {M.}~\bibnamefont {Hemmer}}, \ and\ \bibinfo {author}
  {\bibfnamefont {J.}~\bibnamefont {Biegert}},\ }\href {\doibase
  10.1016/j.pquantelec.2015.07.001} {\bibfield  {journal} {\bibinfo  {journal}
  {Progress in Quantum Electronics}\ }\textbf {\bibinfo {volume} {43}},\
  \bibinfo {pages} {1} (\bibinfo {year} {2015})}\BibitemShut {NoStop}%
\bibitem [{\citenamefont {Katon}(1996)}]{katon_infrared_1996}%
  \BibitemOpen
  \bibfield  {author} {\bibinfo {author} {\bibfnamefont {J.~E.}\ \bibnamefont
  {Katon}},\ }\href {\doibase 10.1016/S0968-4328(96)00045-5} {\bibfield
  {journal} {\bibinfo  {journal} {Micron}\ }\textbf {\bibinfo {volume} {27}},\
  \bibinfo {pages} {303} (\bibinfo {year} {1996})}\BibitemShut {NoStop}%
\bibitem [{\citenamefont {Fernandez}\ \emph {et~al.}(2005)\citenamefont
  {Fernandez}, \citenamefont {Bhargava}, \citenamefont {Hewitt},\ and\
  \citenamefont {Levin}}]{fernandez_infrared_2005}%
  \BibitemOpen
  \bibfield  {author} {\bibinfo {author} {\bibfnamefont {D.~C.}\ \bibnamefont
  {Fernandez}}, \bibinfo {author} {\bibfnamefont {R.}~\bibnamefont {Bhargava}},
  \bibinfo {author} {\bibfnamefont {S.~M.}\ \bibnamefont {Hewitt}}, \ and\
  \bibinfo {author} {\bibfnamefont {I.~W.}\ \bibnamefont {Levin}},\ }\href
  {\doibase 10.1038/nbt1080} {\bibfield  {journal} {\bibinfo  {journal} {Nature
  Biotechnology}\ }\textbf {\bibinfo {volume} {23}},\ \bibinfo {pages} {469}
  (\bibinfo {year} {2005})}\BibitemShut {NoStop}%
\bibitem [{\citenamefont {Bellisola}\ and\ \citenamefont
  {Sorio}(2011)}]{bellisola_infrared_2011}%
  \BibitemOpen
  \bibfield  {author} {\bibinfo {author} {\bibfnamefont {G.}~\bibnamefont
  {Bellisola}}\ and\ \bibinfo {author} {\bibfnamefont {C.}~\bibnamefont
  {Sorio}},\ }\href {https://www.ncbi.nlm.nih.gov/pmc/articles/PMC3236568/}
  {\bibfield  {journal} {\bibinfo  {journal} {American Journal of Cancer
  Research}\ }\textbf {\bibinfo {volume} {2}},\ \bibinfo {pages} {1} (\bibinfo
  {year} {2011})}\BibitemShut {NoStop}%
\bibitem [{\citenamefont {Clemens}\ \emph {et~al.}(2014)\citenamefont
  {Clemens}, \citenamefont {Hands}, \citenamefont {Dorling},\ and\
  \citenamefont {Baker}}]{clemens_vibrational_2014}%
  \BibitemOpen
  \bibfield  {author} {\bibinfo {author} {\bibfnamefont {G.}~\bibnamefont
  {Clemens}}, \bibinfo {author} {\bibfnamefont {J.~R.}\ \bibnamefont {Hands}},
  \bibinfo {author} {\bibfnamefont {K.~M.}\ \bibnamefont {Dorling}}, \ and\
  \bibinfo {author} {\bibfnamefont {M.~J.}\ \bibnamefont {Baker}},\ }\href
  {\doibase 10.1039/C4AN00636D} {\bibfield  {journal} {\bibinfo  {journal}
  {Analyst}\ }\textbf {\bibinfo {volume} {139}},\ \bibinfo {pages} {4411}
  (\bibinfo {year} {2014})}\BibitemShut {NoStop}%
\bibitem [{\citenamefont {Keilmann}\ and\ \citenamefont
  {Hillenbrand}(2004)}]{keilmann_near-field_2004}%
  \BibitemOpen
  \bibfield  {author} {\bibinfo {author} {\bibfnamefont {F.}~\bibnamefont
  {Keilmann}}\ and\ \bibinfo {author} {\bibfnamefont {R.}~\bibnamefont
  {Hillenbrand}},\ }\href {\doibase 10.1098/rsta.2003.1347} {\bibfield
  {journal} {\bibinfo  {journal} {Philosophical Transactions of the Royal
  Society of London A: Mathematical, Physical and Engineering Sciences}\
  }\textbf {\bibinfo {volume} {362}},\ \bibinfo {pages} {787} (\bibinfo {year}
  {2004})}\BibitemShut {NoStop}%
\bibitem [{\citenamefont {Dazzi}\ and\ \citenamefont
  {Prater}(2017)}]{dazzi_afm-ir:_2017}%
  \BibitemOpen
  \bibfield  {author} {\bibinfo {author} {\bibfnamefont {A.}~\bibnamefont
  {Dazzi}}\ and\ \bibinfo {author} {\bibfnamefont {C.~B.}\ \bibnamefont
  {Prater}},\ }\href {\doibase 10.1021/acs.chemrev.6b00448} {\bibfield
  {journal} {\bibinfo  {journal} {Chemical Reviews}\ }\textbf {\bibinfo
  {volume} {117}},\ \bibinfo {pages} {5146} (\bibinfo {year}
  {2017})}\BibitemShut {NoStop}%
\bibitem [{\citenamefont {Bechtel}\ \emph {et~al.}(2014)\citenamefont
  {Bechtel}, \citenamefont {Muller}, \citenamefont {Olmon}, \citenamefont
  {Martin},\ and\ \citenamefont {Raschke}}]{bechtel_ultrabroadband_2014}%
  \BibitemOpen
  \bibfield  {author} {\bibinfo {author} {\bibfnamefont {H.~A.}\ \bibnamefont
  {Bechtel}}, \bibinfo {author} {\bibfnamefont {E.~A.}\ \bibnamefont {Muller}},
  \bibinfo {author} {\bibfnamefont {R.~L.}\ \bibnamefont {Olmon}}, \bibinfo
  {author} {\bibfnamefont {M.~C.}\ \bibnamefont {Martin}}, \ and\ \bibinfo
  {author} {\bibfnamefont {M.~B.}\ \bibnamefont {Raschke}},\ }\href {\doibase
  10.1073/pnas.1400502111} {\bibfield  {journal} {\bibinfo  {journal}
  {Proceedings of the National Academy of Sciences}\ }\textbf {\bibinfo
  {volume} {111}},\ \bibinfo {pages} {7191} (\bibinfo {year}
  {2014})}\BibitemShut {NoStop}%
\bibitem [{\citenamefont {Khatib}\ \emph {et~al.}(2018)\citenamefont {Khatib},
  \citenamefont {Bechtel}, \citenamefont {Martin}, \citenamefont {Raschke},\
  and\ \citenamefont {Carr}}]{khatib_far_2018}%
  \BibitemOpen
  \bibfield  {author} {\bibinfo {author} {\bibfnamefont {O.}~\bibnamefont
  {Khatib}}, \bibinfo {author} {\bibfnamefont {H.~A.}\ \bibnamefont {Bechtel}},
  \bibinfo {author} {\bibfnamefont {M.~C.}\ \bibnamefont {Martin}}, \bibinfo
  {author} {\bibfnamefont {M.~B.}\ \bibnamefont {Raschke}}, \ and\ \bibinfo
  {author} {\bibfnamefont {G.~L.}\ \bibnamefont {Carr}},\ }\href {\doibase
  10.1021/acsphotonics.8b00565} {\bibfield  {journal} {\bibinfo  {journal} {ACS
  Photonics}\ }\textbf {\bibinfo {volume} {5}},\ \bibinfo {pages} {2773}
  (\bibinfo {year} {2018})}\BibitemShut {NoStop}%
\bibitem [{\citenamefont {Bartels}\ \emph {et~al.}(2007)\citenamefont
  {Bartels}, \citenamefont {Cerna}, \citenamefont {Kistner}, \citenamefont
  {Thoma}, \citenamefont {Hudert}, \citenamefont {Janke},\ and\ \citenamefont
  {Dekorsy}}]{bartels_ultrafast_2007}%
  \BibitemOpen
  \bibfield  {author} {\bibinfo {author} {\bibfnamefont {A.}~\bibnamefont
  {Bartels}}, \bibinfo {author} {\bibfnamefont {R.}~\bibnamefont {Cerna}},
  \bibinfo {author} {\bibfnamefont {C.}~\bibnamefont {Kistner}}, \bibinfo
  {author} {\bibfnamefont {A.}~\bibnamefont {Thoma}}, \bibinfo {author}
  {\bibfnamefont {F.}~\bibnamefont {Hudert}}, \bibinfo {author} {\bibfnamefont
  {C.}~\bibnamefont {Janke}}, \ and\ \bibinfo {author} {\bibfnamefont
  {T.}~\bibnamefont {Dekorsy}},\ }\href {\doibase 10.1063/1.2714048} {\bibfield
   {journal} {\bibinfo  {journal} {Review of Scientific Instruments}\ }\textbf
  {\bibinfo {volume} {78}},\ \bibinfo {pages} {035107} (\bibinfo {year}
  {2007})}\BibitemShut {NoStop}%
\bibitem [{\citenamefont {Sell}\ \emph {et~al.}(2008)\citenamefont {Sell},
  \citenamefont {Scheu}, \citenamefont {Leitenstorfer},\ and\ \citenamefont
  {Huber}}]{sell_field-resolved_2008}%
  \BibitemOpen
  \bibfield  {author} {\bibinfo {author} {\bibfnamefont {A.}~\bibnamefont
  {Sell}}, \bibinfo {author} {\bibfnamefont {R.}~\bibnamefont {Scheu}},
  \bibinfo {author} {\bibfnamefont {A.}~\bibnamefont {Leitenstorfer}}, \ and\
  \bibinfo {author} {\bibfnamefont {R.}~\bibnamefont {Huber}},\ }\href
  {\doibase 10.1063/1.3040057} {\bibfield  {journal} {\bibinfo  {journal}
  {Applied Physics Letters}\ }\textbf {\bibinfo {volume} {93}},\ \bibinfo
  {pages} {251107} (\bibinfo {year} {2008})}\BibitemShut {NoStop}%
\bibitem [{\citenamefont {Riek}\ \emph {et~al.}(2015)\citenamefont {Riek},
  \citenamefont {Seletskiy}, \citenamefont {Moskalenko}, \citenamefont
  {Schmidt}, \citenamefont {Krauspe}, \citenamefont {Eckart}, \citenamefont
  {Eggert}, \citenamefont {Burkard},\ and\ \citenamefont
  {Leitenstorfer}}]{riek_direct_2015}%
  \BibitemOpen
  \bibfield  {author} {\bibinfo {author} {\bibfnamefont {C.}~\bibnamefont
  {Riek}}, \bibinfo {author} {\bibfnamefont {D.~V.}\ \bibnamefont {Seletskiy}},
  \bibinfo {author} {\bibfnamefont {A.~S.}\ \bibnamefont {Moskalenko}},
  \bibinfo {author} {\bibfnamefont {J.~F.}\ \bibnamefont {Schmidt}}, \bibinfo
  {author} {\bibfnamefont {P.}~\bibnamefont {Krauspe}}, \bibinfo {author}
  {\bibfnamefont {S.}~\bibnamefont {Eckart}}, \bibinfo {author} {\bibfnamefont
  {S.}~\bibnamefont {Eggert}}, \bibinfo {author} {\bibfnamefont
  {G.}~\bibnamefont {Burkard}}, \ and\ \bibinfo {author} {\bibfnamefont
  {A.}~\bibnamefont {Leitenstorfer}},\ }\href {\doibase
  10.1126/science.aac9788} {\bibfield  {journal} {\bibinfo  {journal}
  {Science}\ }\textbf {\bibinfo {volume} {350}},\ \bibinfo {pages} {420}
  (\bibinfo {year} {2015})}\BibitemShut {NoStop}%
\bibitem [{\citenamefont {Pupeza}\ \emph {et~al.}(2017)\citenamefont {Pupeza},
  \citenamefont {Hubert}, \citenamefont {Schweinberger}, \citenamefont
  {Trubetskov}, \citenamefont {Hussain}, \citenamefont {Vamos}, \citenamefont
  {Pronin}, \citenamefont {Habel}, \citenamefont {Pervak}, \citenamefont
  {Karpowicz}, \citenamefont {Fill}, \citenamefont {Apolonski}, \citenamefont
  {Zigman}, \citenamefont {Azzeer},\ and\ \citenamefont
  {Krausz}}]{pupeza_field-resolved_2017}%
  \BibitemOpen
  \bibfield  {author} {\bibinfo {author} {\bibfnamefont {I.}~\bibnamefont
  {Pupeza}}, \bibinfo {author} {\bibfnamefont {M.}~\bibnamefont {Hubert}},
  \bibinfo {author} {\bibfnamefont {W.}~\bibnamefont {Schweinberger}}, \bibinfo
  {author} {\bibfnamefont {M.}~\bibnamefont {Trubetskov}}, \bibinfo {author}
  {\bibfnamefont {S.~A.}\ \bibnamefont {Hussain}}, \bibinfo {author}
  {\bibfnamefont {L.}~\bibnamefont {Vamos}}, \bibinfo {author} {\bibfnamefont
  {O.}~\bibnamefont {Pronin}}, \bibinfo {author} {\bibfnamefont
  {F.}~\bibnamefont {Habel}}, \bibinfo {author} {\bibfnamefont
  {V.}~\bibnamefont {Pervak}}, \bibinfo {author} {\bibfnamefont
  {N.}~\bibnamefont {Karpowicz}}, \bibinfo {author} {\bibfnamefont
  {E.}~\bibnamefont {Fill}}, \bibinfo {author} {\bibfnamefont {A.}~\bibnamefont
  {Apolonski}}, \bibinfo {author} {\bibfnamefont {M.}~\bibnamefont {Zigman}},
  \bibinfo {author} {\bibfnamefont {A.~M.}\ \bibnamefont {Azzeer}}, \ and\
  \bibinfo {author} {\bibfnamefont {F.}~\bibnamefont {Krausz}},\ }in\ \href
  {\doibase 10.1109/CLEOE-EQEC.2017.8086859} {\emph {\bibinfo {booktitle} {2017
  {Conference} on {Lasers} and {Electro}-{Optics} {Europe} {European} {Quantum}
  {Electronics} {Conference} ({CLEO}/{Europe}-{EQEC})}}}\ (\bibinfo {year}
  {2017})\ pp.\ \bibinfo {pages} {1--1}\BibitemShut {NoStop}%
\bibitem [{\citenamefont {Schliesser}\ \emph {et~al.}(2012)\citenamefont
  {Schliesser}, \citenamefont {Picque},\ and\ \citenamefont
  {Hansch}}]{schliesser_mid-infrared_2012}%
  \BibitemOpen
  \bibfield  {author} {\bibinfo {author} {\bibfnamefont {A.}~\bibnamefont
  {Schliesser}}, \bibinfo {author} {\bibfnamefont {N.}~\bibnamefont {Picque}},
  \ and\ \bibinfo {author} {\bibfnamefont {T.~W.}\ \bibnamefont {Hansch}},\
  }\href {\doibase 10.1038/nphoton.2012.142} {\bibfield  {journal} {\bibinfo
  {journal} {Nat Photon}\ }\textbf {\bibinfo {volume} {6}},\ \bibinfo {pages}
  {440} (\bibinfo {year} {2012})}\BibitemShut {NoStop}%
\bibitem [{\citenamefont {Coddington}\ \emph {et~al.}(2016)\citenamefont
  {Coddington}, \citenamefont {Newbury},\ and\ \citenamefont
  {Swann}}]{coddington_dual-comb_2016}%
  \BibitemOpen
  \bibfield  {author} {\bibinfo {author} {\bibfnamefont {I.}~\bibnamefont
  {Coddington}}, \bibinfo {author} {\bibfnamefont {N.}~\bibnamefont {Newbury}},
  \ and\ \bibinfo {author} {\bibfnamefont {W.}~\bibnamefont {Swann}},\ }\href
  {\doibase 10.1364/OPTICA.3.000414} {\bibfield  {journal} {\bibinfo  {journal}
  {Optica}\ }\textbf {\bibinfo {volume} {3}},\ \bibinfo {pages} {414} (\bibinfo
  {year} {2016})}\BibitemShut {NoStop}%
\bibitem [{\citenamefont {Kara}\ \emph {et~al.}(2017)\citenamefont {Kara},
  \citenamefont {Maidment}, \citenamefont {Gardiner}, \citenamefont
  {Schunemann},\ and\ \citenamefont {Reid}}]{kara_dual-comb_2017}%
  \BibitemOpen
  \bibfield  {author} {\bibinfo {author} {\bibfnamefont {O.}~\bibnamefont
  {Kara}}, \bibinfo {author} {\bibfnamefont {L.}~\bibnamefont {Maidment}},
  \bibinfo {author} {\bibfnamefont {T.}~\bibnamefont {Gardiner}}, \bibinfo
  {author} {\bibfnamefont {P.~G.}\ \bibnamefont {Schunemann}}, \ and\ \bibinfo
  {author} {\bibfnamefont {D.~T.}\ \bibnamefont {Reid}},\ }\href {\doibase
  10.1364/OE.25.032713} {\bibfield  {journal} {\bibinfo  {journal} {Optics
  Express}\ }\textbf {\bibinfo {volume} {25}},\ \bibinfo {pages} {32713}
  (\bibinfo {year} {2017})}\BibitemShut {NoStop}%
\bibitem [{\citenamefont {Maidment}\ \emph {et~al.}(2018)\citenamefont
  {Maidment}, \citenamefont {Kara}, \citenamefont {Schunemann}, \citenamefont
  {Piper}, \citenamefont {McEwan},\ and\ \citenamefont
  {Reid}}]{maidment_long-wave_2018}%
  \BibitemOpen
  \bibfield  {author} {\bibinfo {author} {\bibfnamefont {L.}~\bibnamefont
  {Maidment}}, \bibinfo {author} {\bibfnamefont {O.}~\bibnamefont {Kara}},
  \bibinfo {author} {\bibfnamefont {P.~G.}\ \bibnamefont {Schunemann}},
  \bibinfo {author} {\bibfnamefont {J.}~\bibnamefont {Piper}}, \bibinfo
  {author} {\bibfnamefont {K.}~\bibnamefont {McEwan}}, \ and\ \bibinfo {author}
  {\bibfnamefont {D.~T.}\ \bibnamefont {Reid}},\ }\href {\doibase
  10.1007/s00340-018-7001-2} {\bibfield  {journal} {\bibinfo  {journal}
  {Applied Physics B}\ }\textbf {\bibinfo {volume} {124}},\ \bibinfo {pages}
  {143} (\bibinfo {year} {2018})}\BibitemShut {NoStop}%
\bibitem [{\citenamefont {Muraviev}\ \emph {et~al.}(2018)\citenamefont
  {Muraviev}, \citenamefont {Smolski}, \citenamefont {Loparo},\ and\
  \citenamefont {Vodopyanov}}]{muraviev_massively_2018}%
  \BibitemOpen
  \bibfield  {author} {\bibinfo {author} {\bibfnamefont {A.~V.}\ \bibnamefont
  {Muraviev}}, \bibinfo {author} {\bibfnamefont {V.~O.}\ \bibnamefont
  {Smolski}}, \bibinfo {author} {\bibfnamefont {Z.~E.}\ \bibnamefont {Loparo}},
  \ and\ \bibinfo {author} {\bibfnamefont {K.~L.}\ \bibnamefont {Vodopyanov}},\
  }\href {\doibase 10.1038/s41566-018-0135-2} {\bibfield  {journal} {\bibinfo
  {journal} {Nature Photonics}\ }\textbf {\bibinfo {volume} {12}},\ \bibinfo
  {pages} {209} (\bibinfo {year} {2018})}\BibitemShut {NoStop}%
\bibitem [{\citenamefont {Ycas}\ \emph {et~al.}(2018)\citenamefont {Ycas},
  \citenamefont {Giorgetta}, \citenamefont {Baumann}, \citenamefont
  {Coddington}, \citenamefont {Herman}, \citenamefont {Diddams},\ and\
  \citenamefont {Newbury}}]{ycas_high-coherence_2018}%
  \BibitemOpen
  \bibfield  {author} {\bibinfo {author} {\bibfnamefont {G.}~\bibnamefont
  {Ycas}}, \bibinfo {author} {\bibfnamefont {F.~R.}\ \bibnamefont {Giorgetta}},
  \bibinfo {author} {\bibfnamefont {E.}~\bibnamefont {Baumann}}, \bibinfo
  {author} {\bibfnamefont {I.}~\bibnamefont {Coddington}}, \bibinfo {author}
  {\bibfnamefont {D.}~\bibnamefont {Herman}}, \bibinfo {author} {\bibfnamefont
  {S.~A.}\ \bibnamefont {Diddams}}, \ and\ \bibinfo {author} {\bibfnamefont
  {N.~R.}\ \bibnamefont {Newbury}},\ }\href {\doibase
  10.1038/s41566-018-0114-7} {\bibfield  {journal} {\bibinfo  {journal} {Nature
  Photonics}\ }\textbf {\bibinfo {volume} {12}},\ \bibinfo {pages} {202}
  (\bibinfo {year} {2018})}\BibitemShut {NoStop}%
\bibitem [{\citenamefont {Timmers}\ \emph {et~al.}(2018)\citenamefont
  {Timmers}, \citenamefont {Kowligy}, \citenamefont {Lind}, \citenamefont
  {Cruz}, \citenamefont {Nader}, \citenamefont {Silfies}, \citenamefont {Ycas},
  \citenamefont {Allison}, \citenamefont {Schunemann}, \citenamefont {Papp},\
  and\ \citenamefont {Diddams}}]{timmers_molecular_2018}%
  \BibitemOpen
  \bibfield  {author} {\bibinfo {author} {\bibfnamefont {H.}~\bibnamefont
  {Timmers}}, \bibinfo {author} {\bibfnamefont {A.}~\bibnamefont {Kowligy}},
  \bibinfo {author} {\bibfnamefont {A.}~\bibnamefont {Lind}}, \bibinfo {author}
  {\bibfnamefont {F.~C.}\ \bibnamefont {Cruz}}, \bibinfo {author}
  {\bibfnamefont {N.}~\bibnamefont {Nader}}, \bibinfo {author} {\bibfnamefont
  {M.}~\bibnamefont {Silfies}}, \bibinfo {author} {\bibfnamefont
  {G.}~\bibnamefont {Ycas}}, \bibinfo {author} {\bibfnamefont {T.~K.}\
  \bibnamefont {Allison}}, \bibinfo {author} {\bibfnamefont {P.~G.}\
  \bibnamefont {Schunemann}}, \bibinfo {author} {\bibfnamefont {S.~B.}\
  \bibnamefont {Papp}}, \ and\ \bibinfo {author} {\bibfnamefont {S.~A.}\
  \bibnamefont {Diddams}},\ }\href {\doibase 10.1364/OPTICA.5.000727}
  {\bibfield  {journal} {\bibinfo  {journal} {Optica}\ }\textbf {\bibinfo
  {volume} {5}},\ \bibinfo {pages} {727} (\bibinfo {year} {2018})}\BibitemShut
  {NoStop}%
\bibitem [{\citenamefont {Newbury}\ \emph {et~al.}(2010)\citenamefont
  {Newbury}, \citenamefont {Coddington},\ and\ \citenamefont
  {Swann}}]{newbury_sensitivity_2010}%
  \BibitemOpen
  \bibfield  {author} {\bibinfo {author} {\bibfnamefont {N.~R.}\ \bibnamefont
  {Newbury}}, \bibinfo {author} {\bibfnamefont {I.}~\bibnamefont {Coddington}},
  \ and\ \bibinfo {author} {\bibfnamefont {W.}~\bibnamefont {Swann}},\ }\href
  {\doibase 10.1364/OE.18.007929} {\bibfield  {journal} {\bibinfo  {journal}
  {Optics Express}\ }\textbf {\bibinfo {volume} {18}},\ \bibinfo {pages} {7929}
  (\bibinfo {year} {2010})}\BibitemShut {NoStop}%
\bibitem [{\citenamefont {Baltu{\v s}ka}\ \emph {et~al.}(2002)\citenamefont
  {Baltu{\v s}ka}, \citenamefont {Fuji},\ and\ \citenamefont
  {Kobayashi}}]{baltuska_controlling_2002}%
  \BibitemOpen
  \bibfield  {author} {\bibinfo {author} {\bibfnamefont {A.}~\bibnamefont
  {Baltu{\v s}ka}}, \bibinfo {author} {\bibfnamefont {T.}~\bibnamefont {Fuji}},
  \ and\ \bibinfo {author} {\bibfnamefont {T.}~\bibnamefont {Kobayashi}},\
  }\href {\doibase 10.1103/PhysRevLett.88.133901} {\bibfield  {journal}
  {\bibinfo  {journal} {Physical Review Letters}\ }\textbf {\bibinfo {volume}
  {88}},\ \bibinfo {pages} {133901} (\bibinfo {year} {2002})}\BibitemShut
  {NoStop}%
\bibitem [{\citenamefont {Wu}\ and\ \citenamefont
  {Zhang}(1995)}]{wu_freespace_1995}%
  \BibitemOpen
  \bibfield  {author} {\bibinfo {author} {\bibfnamefont {Q.}~\bibnamefont
  {Wu}}\ and\ \bibinfo {author} {\bibfnamefont {X.-C.}\ \bibnamefont {Zhang}},\
  }\href {\doibase 10.1063/1.114909} {\bibfield  {journal} {\bibinfo  {journal}
  {Applied Physics Letters}\ }\textbf {\bibinfo {volume} {67}},\ \bibinfo
  {pages} {3523} (\bibinfo {year} {1995})}\BibitemShut {NoStop}%
\bibitem [{\citenamefont {Pupeza}\ \emph {et~al.}(2015)\citenamefont {Pupeza},
  \citenamefont {S{\'a}nchez}, \citenamefont {Zhang}, \citenamefont
  {Lilienfein}, \citenamefont {Seidel}, \citenamefont {Karpowicz},
  \citenamefont {Paasch-Colberg}, \citenamefont {Znakovskaya}, \citenamefont
  {Pescher}, \citenamefont {Schweinberger}, \citenamefont {Pervak},
  \citenamefont {Fill}, \citenamefont {Pronin}, \citenamefont {Wei},
  \citenamefont {Krausz}, \citenamefont {Apolonski},\ and\ \citenamefont
  {Biegert}}]{pupeza_high-power_2015}%
  \BibitemOpen
  \bibfield  {author} {\bibinfo {author} {\bibfnamefont {I.}~\bibnamefont
  {Pupeza}}, \bibinfo {author} {\bibfnamefont {D.}~\bibnamefont {S{\'a}nchez}},
  \bibinfo {author} {\bibfnamefont {J.}~\bibnamefont {Zhang}}, \bibinfo
  {author} {\bibfnamefont {N.}~\bibnamefont {Lilienfein}}, \bibinfo {author}
  {\bibfnamefont {M.}~\bibnamefont {Seidel}}, \bibinfo {author} {\bibfnamefont
  {N.}~\bibnamefont {Karpowicz}}, \bibinfo {author} {\bibfnamefont
  {T.}~\bibnamefont {Paasch-Colberg}}, \bibinfo {author} {\bibfnamefont
  {I.}~\bibnamefont {Znakovskaya}}, \bibinfo {author} {\bibfnamefont
  {M.}~\bibnamefont {Pescher}}, \bibinfo {author} {\bibfnamefont
  {W.}~\bibnamefont {Schweinberger}}, \bibinfo {author} {\bibfnamefont
  {V.}~\bibnamefont {Pervak}}, \bibinfo {author} {\bibfnamefont
  {E.}~\bibnamefont {Fill}}, \bibinfo {author} {\bibfnamefont {O.}~\bibnamefont
  {Pronin}}, \bibinfo {author} {\bibfnamefont {Z.}~\bibnamefont {Wei}},
  \bibinfo {author} {\bibfnamefont {F.}~\bibnamefont {Krausz}}, \bibinfo
  {author} {\bibfnamefont {A.}~\bibnamefont {Apolonski}}, \ and\ \bibinfo
  {author} {\bibfnamefont {J.}~\bibnamefont {Biegert}},\ }\href {\doibase
  10.1038/nphoton.2015.179} {\bibfield  {journal} {\bibinfo  {journal} {Nature
  Photonics}\ }\textbf {\bibinfo {volume} {9}},\ \bibinfo {pages} {721}
  (\bibinfo {year} {2015})}\BibitemShut {NoStop}%
\bibitem [{\citenamefont {Huber}\ \emph {et~al.}(2017)\citenamefont {Huber},
  \citenamefont {Schweinberger}, \citenamefont {Trubetskov}, \citenamefont
  {Hussain}, \citenamefont {Pronin}, \citenamefont {Vamos}, \citenamefont
  {Fill}, \citenamefont {Apolonski}, \citenamefont {Zigman}, \citenamefont
  {Krausz},\ and\ \citenamefont {Pupeza}}]{KrauszConf2}%
  \BibitemOpen
  \bibfield  {author} {\bibinfo {author} {\bibfnamefont {M.}~\bibnamefont
  {Huber}}, \bibinfo {author} {\bibfnamefont {W.}~\bibnamefont
  {Schweinberger}}, \bibinfo {author} {\bibfnamefont {M.}~\bibnamefont
  {Trubetskov}}, \bibinfo {author} {\bibfnamefont {S.~A.}\ \bibnamefont
  {Hussain}}, \bibinfo {author} {\bibfnamefont {O.}~\bibnamefont {Pronin}},
  \bibinfo {author} {\bibfnamefont {L.}~\bibnamefont {Vamos}}, \bibinfo
  {author} {\bibfnamefont {E.}~\bibnamefont {Fill}}, \bibinfo {author}
  {\bibfnamefont {A.}~\bibnamefont {Apolonski}}, \bibinfo {author}
  {\bibfnamefont {M.}~\bibnamefont {Zigman}}, \bibinfo {author} {\bibfnamefont
  {F.}~\bibnamefont {Krausz}}, \ and\ \bibinfo {author} {\bibfnamefont
  {I.}~\bibnamefont {Pupeza}},\ }in\ \href {\doibase
  10.1109/CLEOE-EQEC.2017.8086921} {\emph {\bibinfo {booktitle} {2017
  Conference on Lasers and Electro-Optics Europe European Quantum Electronics
  Conference (CLEO/Europe-EQEC)}}}\ (\bibinfo {year} {2017})\ pp.\ \bibinfo
  {pages} {1--1}\BibitemShut {NoStop}%
\bibitem [{\citenamefont {Coddington}\ \emph
  {et~al.}(2009{\natexlab{a}})\citenamefont {Coddington}, \citenamefont
  {Swann}, \citenamefont {Nenadovic},\ and\ \citenamefont
  {Newbury}}]{coddington_rapid_2009}%
  \BibitemOpen
  \bibfield  {author} {\bibinfo {author} {\bibfnamefont {I.}~\bibnamefont
  {Coddington}}, \bibinfo {author} {\bibfnamefont {W.~C.}\ \bibnamefont
  {Swann}}, \bibinfo {author} {\bibfnamefont {L.}~\bibnamefont {Nenadovic}}, \
  and\ \bibinfo {author} {\bibfnamefont {N.~R.}\ \bibnamefont {Newbury}},\
  }\href {\doibase 10.1038/nphoton.2009.94} {\bibfield  {journal} {\bibinfo
  {journal} {Nature Photonics}\ }\textbf {\bibinfo {volume} {3}},\ \bibinfo
  {pages} {351} (\bibinfo {year} {2009}{\natexlab{a}})}\BibitemShut {NoStop}%
\bibitem [{\citenamefont {Coddington}\ \emph
  {et~al.}(2009{\natexlab{b}})\citenamefont {Coddington}, \citenamefont
  {Swann},\ and\ \citenamefont {Newbury}}]{coddington_coherent_2009}%
  \BibitemOpen
  \bibfield  {author} {\bibinfo {author} {\bibfnamefont {I.}~\bibnamefont
  {Coddington}}, \bibinfo {author} {\bibfnamefont {W.~C.}\ \bibnamefont
  {Swann}}, \ and\ \bibinfo {author} {\bibfnamefont {N.~R.}\ \bibnamefont
  {Newbury}},\ }\href {\doibase 10.1364/OL.34.002153} {\bibfield  {journal}
  {\bibinfo  {journal} {Optics Letters}\ }\textbf {\bibinfo {volume} {34}},\
  \bibinfo {pages} {2153} (\bibinfo {year} {2009}{\natexlab{b}})}\BibitemShut
  {NoStop}%
\bibitem [{\citenamefont {Coddington}\ \emph {et~al.}(2010)\citenamefont
  {Coddington}, \citenamefont {Swann},\ and\ \citenamefont
  {Newbury}}]{coddington_time-domain_2010}%
  \BibitemOpen
  \bibfield  {author} {\bibinfo {author} {\bibfnamefont {I.}~\bibnamefont
  {Coddington}}, \bibinfo {author} {\bibfnamefont {W.~C.}\ \bibnamefont
  {Swann}}, \ and\ \bibinfo {author} {\bibfnamefont {N.~R.}\ \bibnamefont
  {Newbury}},\ }\href {\doibase 10.1364/OL.35.001395} {\bibfield  {journal}
  {\bibinfo  {journal} {Optics Letters}\ }\textbf {\bibinfo {volume} {35}},\
  \bibinfo {pages} {1395} (\bibinfo {year} {2010})}\BibitemShut {NoStop}%
\bibitem [{\citenamefont {Stead}\ \emph {et~al.}(2012)\citenamefont {Stead},
  \citenamefont {Mills},\ and\ \citenamefont {Jones}}]{stead_method_2012}%
  \BibitemOpen
  \bibfield  {author} {\bibinfo {author} {\bibfnamefont {R.~A.}\ \bibnamefont
  {Stead}}, \bibinfo {author} {\bibfnamefont {A.~K.}\ \bibnamefont {Mills}}, \
  and\ \bibinfo {author} {\bibfnamefont {D.~J.}\ \bibnamefont {Jones}},\ }\href
  {\doibase 10.1364/JOSAB.29.002861} {\bibfield  {journal} {\bibinfo  {journal}
  {JOSA B}\ }\textbf {\bibinfo {volume} {29}},\ \bibinfo {pages} {2861}
  (\bibinfo {year} {2012})}\BibitemShut {NoStop}%
\bibitem [{\citenamefont {Ideguchi}\ \emph {et~al.}(2013)\citenamefont
  {Ideguchi}, \citenamefont {Holzner}, \citenamefont {Bernhardt}, \citenamefont
  {Guelachvili}, \citenamefont {Picqu{\'e}},\ and\ \citenamefont
  {H{\"a}nsch}}]{ideguchi_coherent_2013}%
  \BibitemOpen
  \bibfield  {author} {\bibinfo {author} {\bibfnamefont {T.}~\bibnamefont
  {Ideguchi}}, \bibinfo {author} {\bibfnamefont {S.}~\bibnamefont {Holzner}},
  \bibinfo {author} {\bibfnamefont {B.}~\bibnamefont {Bernhardt}}, \bibinfo
  {author} {\bibfnamefont {G.}~\bibnamefont {Guelachvili}}, \bibinfo {author}
  {\bibfnamefont {N.}~\bibnamefont {Picqu{\'e}}}, \ and\ \bibinfo {author}
  {\bibfnamefont {T.~W.}\ \bibnamefont {H{\"a}nsch}},\ }\href {\doibase
  10.1038/nature12607} {\bibfield  {journal} {\bibinfo  {journal} {Nature}\
  }\textbf {\bibinfo {volume} {502}},\ \bibinfo {pages} {355} (\bibinfo {year}
  {2013})}\BibitemShut {NoStop}%
\bibitem [{\citenamefont {Timmer}\ \emph {et~al.}(2005)\citenamefont {Timmer},
  \citenamefont {Olthuis},\ and\ \citenamefont {van~den
  Berg}}]{timmer_ammonia_2005}%
  \BibitemOpen
  \bibfield  {author} {\bibinfo {author} {\bibfnamefont {B.}~\bibnamefont
  {Timmer}}, \bibinfo {author} {\bibfnamefont {W.}~\bibnamefont {Olthuis}}, \
  and\ \bibinfo {author} {\bibfnamefont {A.}~\bibnamefont {van~den Berg}},\
  }\href {\doibase 10.1016/j.snb.2004.11.054} {\bibfield  {journal} {\bibinfo
  {journal} {Sensors and Actuators B: Chemical}\ }\textbf {\bibinfo {volume}
  {107}},\ \bibinfo {pages} {666} (\bibinfo {year} {2005})}\BibitemShut
  {NoStop}%
\bibitem [{\citenamefont {Schliesser}\ \emph {et~al.}(2005)\citenamefont
  {Schliesser}, \citenamefont {Brehm}, \citenamefont {Keilmann},\ and\
  \citenamefont {van~der Weide}}]{Schliesser:05}%
  \BibitemOpen
  \bibfield  {author} {\bibinfo {author} {\bibfnamefont {A.}~\bibnamefont
  {Schliesser}}, \bibinfo {author} {\bibfnamefont {M.}~\bibnamefont {Brehm}},
  \bibinfo {author} {\bibfnamefont {F.}~\bibnamefont {Keilmann}}, \ and\
  \bibinfo {author} {\bibfnamefont {D.~W.}\ \bibnamefont {van~der Weide}},\
  }\href {\doibase 10.1364/OPEX.13.009029} {\bibfield  {journal} {\bibinfo
  {journal} {Opt. Express}\ }\textbf {\bibinfo {volume} {13}},\ \bibinfo
  {pages} {9029} (\bibinfo {year} {2005})}\BibitemShut {NoStop}%
\bibitem [{\citenamefont {Kirkby}\ \emph {et~al.}(2011)\citenamefont {Kirkby},
  \citenamefont {Curtius}, \citenamefont {Almeida}, \citenamefont {Dunne},
  \citenamefont {Duplissy}, \citenamefont {Ehrhart}, \citenamefont {Franchin},
  \citenamefont {Gagn{\'e}}, \citenamefont {Ickes}, \citenamefont {K{\"u}rten},
  \citenamefont {Kupc}, \citenamefont {Metzger}, \citenamefont {Riccobono},
  \citenamefont {Rondo}, \citenamefont {Schobesberger}, \citenamefont
  {Tsagkogeorgas}, \citenamefont {Wimmer}, \citenamefont {Amorim},
  \citenamefont {Bianchi}, \citenamefont {Breitenlechner}, \citenamefont
  {David}, \citenamefont {Dommen}, \citenamefont {Downard}, \citenamefont
  {Ehn}, \citenamefont {Flagan}, \citenamefont {Haider}, \citenamefont
  {Hansel}, \citenamefont {Hauser}, \citenamefont {Jud}, \citenamefont
  {Junninen}, \citenamefont {Kreissl}, \citenamefont {Kvashin}, \citenamefont
  {Laaksonen}, \citenamefont {Lehtipalo}, \citenamefont {Lima}, \citenamefont
  {Lovejoy}, \citenamefont {Makhmutov}, \citenamefont {Mathot}, \citenamefont
  {Mikkil{\"a}}, \citenamefont {Minginette}, \citenamefont {Mogo},
  \citenamefont {Nieminen}, \citenamefont {Onnela}, \citenamefont {Pereira},
  \citenamefont {Pet{\"a}j{\"a}}, \citenamefont {Schnitzhofer}, \citenamefont
  {Seinfeld}, \citenamefont {Sipil{\"a}}, \citenamefont {Stozhkov},
  \citenamefont {Stratmann}, \citenamefont {Tom{\'e}}, \citenamefont
  {Vanhanen}, \citenamefont {Viisanen}, \citenamefont {Vrtala}, \citenamefont
  {Wagner}, \citenamefont {Walther}, \citenamefont {Weingartner}, \citenamefont
  {Wex}, \citenamefont {Winkler}, \citenamefont {Carslaw}, \citenamefont
  {Worsnop}, \citenamefont {Baltensperger},\ and\ \citenamefont
  {Kulmala}}]{kirkby_role_2011}%
  \BibitemOpen
  \bibfield  {author} {\bibinfo {author} {\bibfnamefont {J.}~\bibnamefont
  {Kirkby}}, \bibinfo {author} {\bibfnamefont {J.}~\bibnamefont {Curtius}},
  \bibinfo {author} {\bibfnamefont {J.}~\bibnamefont {Almeida}}, \bibinfo
  {author} {\bibfnamefont {E.}~\bibnamefont {Dunne}}, \bibinfo {author}
  {\bibfnamefont {J.}~\bibnamefont {Duplissy}}, \bibinfo {author}
  {\bibfnamefont {S.}~\bibnamefont {Ehrhart}}, \bibinfo {author} {\bibfnamefont
  {A.}~\bibnamefont {Franchin}}, \bibinfo {author} {\bibfnamefont
  {S.}~\bibnamefont {Gagn{\'e}}}, \bibinfo {author} {\bibfnamefont
  {L.}~\bibnamefont {Ickes}}, \bibinfo {author} {\bibfnamefont
  {A.}~\bibnamefont {K{\"u}rten}}, \bibinfo {author} {\bibfnamefont
  {A.}~\bibnamefont {Kupc}}, \bibinfo {author} {\bibfnamefont {A.}~\bibnamefont
  {Metzger}}, \bibinfo {author} {\bibfnamefont {F.}~\bibnamefont {Riccobono}},
  \bibinfo {author} {\bibfnamefont {L.}~\bibnamefont {Rondo}}, \bibinfo
  {author} {\bibfnamefont {S.}~\bibnamefont {Schobesberger}}, \bibinfo {author}
  {\bibfnamefont {G.}~\bibnamefont {Tsagkogeorgas}}, \bibinfo {author}
  {\bibfnamefont {D.}~\bibnamefont {Wimmer}}, \bibinfo {author} {\bibfnamefont
  {A.}~\bibnamefont {Amorim}}, \bibinfo {author} {\bibfnamefont
  {F.}~\bibnamefont {Bianchi}}, \bibinfo {author} {\bibfnamefont
  {M.}~\bibnamefont {Breitenlechner}}, \bibinfo {author} {\bibfnamefont
  {A.}~\bibnamefont {David}}, \bibinfo {author} {\bibfnamefont
  {J.}~\bibnamefont {Dommen}}, \bibinfo {author} {\bibfnamefont
  {A.}~\bibnamefont {Downard}}, \bibinfo {author} {\bibfnamefont
  {M.}~\bibnamefont {Ehn}}, \bibinfo {author} {\bibfnamefont {R.~C.}\
  \bibnamefont {Flagan}}, \bibinfo {author} {\bibfnamefont {S.}~\bibnamefont
  {Haider}}, \bibinfo {author} {\bibfnamefont {A.}~\bibnamefont {Hansel}},
  \bibinfo {author} {\bibfnamefont {D.}~\bibnamefont {Hauser}}, \bibinfo
  {author} {\bibfnamefont {W.}~\bibnamefont {Jud}}, \bibinfo {author}
  {\bibfnamefont {H.}~\bibnamefont {Junninen}}, \bibinfo {author}
  {\bibfnamefont {F.}~\bibnamefont {Kreissl}}, \bibinfo {author} {\bibfnamefont
  {A.}~\bibnamefont {Kvashin}}, \bibinfo {author} {\bibfnamefont
  {A.}~\bibnamefont {Laaksonen}}, \bibinfo {author} {\bibfnamefont
  {K.}~\bibnamefont {Lehtipalo}}, \bibinfo {author} {\bibfnamefont
  {J.}~\bibnamefont {Lima}}, \bibinfo {author} {\bibfnamefont {E.~R.}\
  \bibnamefont {Lovejoy}}, \bibinfo {author} {\bibfnamefont {V.}~\bibnamefont
  {Makhmutov}}, \bibinfo {author} {\bibfnamefont {S.}~\bibnamefont {Mathot}},
  \bibinfo {author} {\bibfnamefont {J.}~\bibnamefont {Mikkil{\"a}}}, \bibinfo
  {author} {\bibfnamefont {P.}~\bibnamefont {Minginette}}, \bibinfo {author}
  {\bibfnamefont {S.}~\bibnamefont {Mogo}}, \bibinfo {author} {\bibfnamefont
  {T.}~\bibnamefont {Nieminen}}, \bibinfo {author} {\bibfnamefont
  {A.}~\bibnamefont {Onnela}}, \bibinfo {author} {\bibfnamefont
  {P.}~\bibnamefont {Pereira}}, \bibinfo {author} {\bibfnamefont
  {T.}~\bibnamefont {Pet{\"a}j{\"a}}}, \bibinfo {author} {\bibfnamefont
  {R.}~\bibnamefont {Schnitzhofer}}, \bibinfo {author} {\bibfnamefont {J.~H.}\
  \bibnamefont {Seinfeld}}, \bibinfo {author} {\bibfnamefont {M.}~\bibnamefont
  {Sipil{\"a}}}, \bibinfo {author} {\bibfnamefont {Y.}~\bibnamefont
  {Stozhkov}}, \bibinfo {author} {\bibfnamefont {F.}~\bibnamefont {Stratmann}},
  \bibinfo {author} {\bibfnamefont {A.}~\bibnamefont {Tom{\'e}}}, \bibinfo
  {author} {\bibfnamefont {J.}~\bibnamefont {Vanhanen}}, \bibinfo {author}
  {\bibfnamefont {Y.}~\bibnamefont {Viisanen}}, \bibinfo {author}
  {\bibfnamefont {A.}~\bibnamefont {Vrtala}}, \bibinfo {author} {\bibfnamefont
  {P.~E.}\ \bibnamefont {Wagner}}, \bibinfo {author} {\bibfnamefont
  {H.}~\bibnamefont {Walther}}, \bibinfo {author} {\bibfnamefont
  {E.}~\bibnamefont {Weingartner}}, \bibinfo {author} {\bibfnamefont
  {H.}~\bibnamefont {Wex}}, \bibinfo {author} {\bibfnamefont {P.~M.}\
  \bibnamefont {Winkler}}, \bibinfo {author} {\bibfnamefont {K.~S.}\
  \bibnamefont {Carslaw}}, \bibinfo {author} {\bibfnamefont {D.~R.}\
  \bibnamefont {Worsnop}}, \bibinfo {author} {\bibfnamefont {U.}~\bibnamefont
  {Baltensperger}}, \ and\ \bibinfo {author} {\bibfnamefont {M.}~\bibnamefont
  {Kulmala}},\ }\href {\doibase 10.1038/nature10343} {\bibfield  {journal}
  {\bibinfo  {journal} {Nature}\ }\textbf {\bibinfo {volume} {476}},\ \bibinfo
  {pages} {429} (\bibinfo {year} {2011})}\BibitemShut {NoStop}%
\bibitem [{\citenamefont {Cossel}\ \emph {et~al.}(2017)\citenamefont {Cossel},
  \citenamefont {Waxman}, \citenamefont {Finneran}, \citenamefont {Blake},
  \citenamefont {Ye},\ and\ \citenamefont {Newbury}}]{cossel_gas-phase_2017}%
  \BibitemOpen
  \bibfield  {author} {\bibinfo {author} {\bibfnamefont {K.~C.}\ \bibnamefont
  {Cossel}}, \bibinfo {author} {\bibfnamefont {E.~M.}\ \bibnamefont {Waxman}},
  \bibinfo {author} {\bibfnamefont {I.~A.}\ \bibnamefont {Finneran}}, \bibinfo
  {author} {\bibfnamefont {G.~A.}\ \bibnamefont {Blake}}, \bibinfo {author}
  {\bibfnamefont {J.}~\bibnamefont {Ye}}, \ and\ \bibinfo {author}
  {\bibfnamefont {N.~R.}\ \bibnamefont {Newbury}},\ }\href {\doibase
  10.1364/JOSAB.34.000104} {\bibfield  {journal} {\bibinfo  {journal} {JOSA B}\
  }\textbf {\bibinfo {volume} {34}},\ \bibinfo {pages} {104} (\bibinfo {year}
  {2017})}\BibitemShut {NoStop}%
\bibitem [{\citenamefont {Spaun}\ \emph {et~al.}(2016)\citenamefont {Spaun},
  \citenamefont {Changala}, \citenamefont {Patterson}, \citenamefont {Bjork},
  \citenamefont {Heckl}, \citenamefont {Doyle},\ and\ \citenamefont
  {Ye}}]{spaun_continuous_2016}%
  \BibitemOpen
  \bibfield  {author} {\bibinfo {author} {\bibfnamefont {B.}~\bibnamefont
  {Spaun}}, \bibinfo {author} {\bibfnamefont {P.~B.}\ \bibnamefont {Changala}},
  \bibinfo {author} {\bibfnamefont {D.}~\bibnamefont {Patterson}}, \bibinfo
  {author} {\bibfnamefont {B.~J.}\ \bibnamefont {Bjork}}, \bibinfo {author}
  {\bibfnamefont {O.~H.}\ \bibnamefont {Heckl}}, \bibinfo {author}
  {\bibfnamefont {J.~M.}\ \bibnamefont {Doyle}}, \ and\ \bibinfo {author}
  {\bibfnamefont {J.}~\bibnamefont {Ye}},\ }\href {\doibase
  10.1038/nature17440} {\bibfield  {journal} {\bibinfo  {journal} {Nature}\
  }\textbf {\bibinfo {volume} {533}},\ \bibinfo {pages} {517} (\bibinfo {year}
  {2016})}\BibitemShut {NoStop}%
\bibitem [{\citenamefont {Fleisher}\ \emph {et~al.}(2014)\citenamefont
  {Fleisher}, \citenamefont {Bjork}, \citenamefont {Bui}, \citenamefont
  {Cossel}, \citenamefont {Okumura},\ and\ \citenamefont
  {Ye}}]{fleisher_mid-infrared_2014}%
  \BibitemOpen
  \bibfield  {author} {\bibinfo {author} {\bibfnamefont {A.~J.}\ \bibnamefont
  {Fleisher}}, \bibinfo {author} {\bibfnamefont {B.~J.}\ \bibnamefont {Bjork}},
  \bibinfo {author} {\bibfnamefont {T.~Q.}\ \bibnamefont {Bui}}, \bibinfo
  {author} {\bibfnamefont {K.~C.}\ \bibnamefont {Cossel}}, \bibinfo {author}
  {\bibfnamefont {M.}~\bibnamefont {Okumura}}, \ and\ \bibinfo {author}
  {\bibfnamefont {J.}~\bibnamefont {Ye}},\ }\href {\doibase 10.1021/jz5008559}
  {\bibfield  {journal} {\bibinfo  {journal} {The Journal of Physical Chemistry
  Letters}\ }\textbf {\bibinfo {volume} {5}},\ \bibinfo {pages} {2241}
  (\bibinfo {year} {2014})}\BibitemShut {NoStop}%
\bibitem [{\citenamefont {Dong}\ \emph {et~al.}(2018)\citenamefont {Dong},
  \citenamefont {Liang}, \citenamefont {Yan}, \citenamefont {Markey},
  \citenamefont {Mirokhin}, \citenamefont {Tchekhovskoi}, \citenamefont
  {Bukhari},\ and\ \citenamefont {Stein}}]{dong_nistmab_2018}%
  \BibitemOpen
  \bibfield  {author} {\bibinfo {author} {\bibfnamefont {Q.}~\bibnamefont
  {Dong}}, \bibinfo {author} {\bibfnamefont {Y.}~\bibnamefont {Liang}},
  \bibinfo {author} {\bibfnamefont {X.}~\bibnamefont {Yan}}, \bibinfo {author}
  {\bibfnamefont {S.~P.}\ \bibnamefont {Markey}}, \bibinfo {author}
  {\bibfnamefont {Y.~A.}\ \bibnamefont {Mirokhin}}, \bibinfo {author}
  {\bibfnamefont {D.~V.}\ \bibnamefont {Tchekhovskoi}}, \bibinfo {author}
  {\bibfnamefont {T.~H.}\ \bibnamefont {Bukhari}}, \ and\ \bibinfo {author}
  {\bibfnamefont {S.~E.}\ \bibnamefont {Stein}},\ }\href {\doibase
  10.1080/19420862.2018.1436921} {\bibfield  {journal} {\bibinfo  {journal}
  {mAbs}\ }\textbf {\bibinfo {volume} {10}},\ \bibinfo {pages} {354} (\bibinfo
  {year} {2018})}\BibitemShut {NoStop}%
\bibitem [{\citenamefont {Baker}\ \emph {et~al.}(2008)\citenamefont {Baker},
  \citenamefont {Gazi}, \citenamefont {Brown}, \citenamefont {Shanks},
  \citenamefont {Gardner},\ and\ \citenamefont
  {Clarke}}]{baker_ftir-based_2008}%
  \BibitemOpen
  \bibfield  {author} {\bibinfo {author} {\bibfnamefont {M.~J.}\ \bibnamefont
  {Baker}}, \bibinfo {author} {\bibfnamefont {E.}~\bibnamefont {Gazi}},
  \bibinfo {author} {\bibfnamefont {M.~D.}\ \bibnamefont {Brown}}, \bibinfo
  {author} {\bibfnamefont {J.~H.}\ \bibnamefont {Shanks}}, \bibinfo {author}
  {\bibfnamefont {P.}~\bibnamefont {Gardner}}, \ and\ \bibinfo {author}
  {\bibfnamefont {N.~W.}\ \bibnamefont {Clarke}},\ }\href {\doibase
  10.1038/sj.bjc.6604753} {\bibfield  {journal} {\bibinfo  {journal} {British
  Journal of Cancer}\ }\textbf {\bibinfo {volume} {99}},\ \bibinfo {pages}
  {1859} (\bibinfo {year} {2008})}\BibitemShut {NoStop}%
\bibitem [{\citenamefont {Baker}\ \emph {et~al.}(2014)\citenamefont {Baker},
  \citenamefont {Trevisan}, \citenamefont {Bassan}, \citenamefont {Bhargava},
  \citenamefont {Butler}, \citenamefont {Dorling}, \citenamefont {Fielden},
  \citenamefont {Fogarty}, \citenamefont {Fullwood}, \citenamefont {Heys},
  \citenamefont {Hughes}, \citenamefont {Lasch}, \citenamefont {Martin-Hirsch},
  \citenamefont {Obinaju}, \citenamefont {Sockalingum}, \citenamefont
  {Sul{\'e}-Suso}, \citenamefont {Strong}, \citenamefont {Walsh}, \citenamefont
  {Wood}, \citenamefont {Gardner},\ and\ \citenamefont
  {Martin}}]{baker_using_2014}%
  \BibitemOpen
  \bibfield  {author} {\bibinfo {author} {\bibfnamefont {M.~J.}\ \bibnamefont
  {Baker}}, \bibinfo {author} {\bibfnamefont {J.}~\bibnamefont {Trevisan}},
  \bibinfo {author} {\bibfnamefont {P.}~\bibnamefont {Bassan}}, \bibinfo
  {author} {\bibfnamefont {R.}~\bibnamefont {Bhargava}}, \bibinfo {author}
  {\bibfnamefont {H.~J.}\ \bibnamefont {Butler}}, \bibinfo {author}
  {\bibfnamefont {K.~M.}\ \bibnamefont {Dorling}}, \bibinfo {author}
  {\bibfnamefont {P.~R.}\ \bibnamefont {Fielden}}, \bibinfo {author}
  {\bibfnamefont {S.~W.}\ \bibnamefont {Fogarty}}, \bibinfo {author}
  {\bibfnamefont {N.~J.}\ \bibnamefont {Fullwood}}, \bibinfo {author}
  {\bibfnamefont {K.~A.}\ \bibnamefont {Heys}}, \bibinfo {author}
  {\bibfnamefont {C.}~\bibnamefont {Hughes}}, \bibinfo {author} {\bibfnamefont
  {P.}~\bibnamefont {Lasch}}, \bibinfo {author} {\bibfnamefont {P.~L.}\
  \bibnamefont {Martin-Hirsch}}, \bibinfo {author} {\bibfnamefont
  {B.}~\bibnamefont {Obinaju}}, \bibinfo {author} {\bibfnamefont {G.~D.}\
  \bibnamefont {Sockalingum}}, \bibinfo {author} {\bibfnamefont
  {J.}~\bibnamefont {Sul{\'e}-Suso}}, \bibinfo {author} {\bibfnamefont {R.~J.}\
  \bibnamefont {Strong}}, \bibinfo {author} {\bibfnamefont {M.~J.}\
  \bibnamefont {Walsh}}, \bibinfo {author} {\bibfnamefont {B.~R.}\ \bibnamefont
  {Wood}}, \bibinfo {author} {\bibfnamefont {P.}~\bibnamefont {Gardner}}, \
  and\ \bibinfo {author} {\bibfnamefont {F.~L.}\ \bibnamefont {Martin}},\
  }\href {\doibase 10.1038/nprot.2014.110} {\bibfield  {journal} {\bibinfo
  {journal} {Nature Protocols}\ }\textbf {\bibinfo {volume} {9}},\ \bibinfo
  {pages} {1771} (\bibinfo {year} {2014})}\BibitemShut {NoStop}%
\bibitem [{\citenamefont {Gokarn}\ \emph {et~al.}(2015)\citenamefont {Gokarn},
  \citenamefont {Agarwal}, \citenamefont {Arthur}, \citenamefont {Bepperling},
  \citenamefont {Day}, \citenamefont {Filoti}, \citenamefont {Greene},
  \citenamefont {Hayes}, \citenamefont {Kroe-Barrett}, \citenamefont {Laue},
  \citenamefont {Lin}, \citenamefont {McGarry}, \citenamefont {Razinkov},
  \citenamefont {Singh}, \citenamefont {Taing}, \citenamefont {Venkataramani},
  \citenamefont {Weiss}, \citenamefont {Yang},\ and\ \citenamefont
  {Zarraga}}]{gokarn_biophysical_2015}%
  \BibitemOpen
  \bibfield  {author} {\bibinfo {author} {\bibfnamefont {Y.}~\bibnamefont
  {Gokarn}}, \bibinfo {author} {\bibfnamefont {S.}~\bibnamefont {Agarwal}},
  \bibinfo {author} {\bibfnamefont {K.}~\bibnamefont {Arthur}}, \bibinfo
  {author} {\bibfnamefont {A.}~\bibnamefont {Bepperling}}, \bibinfo {author}
  {\bibfnamefont {E.~S.}\ \bibnamefont {Day}}, \bibinfo {author} {\bibfnamefont
  {D.}~\bibnamefont {Filoti}}, \bibinfo {author} {\bibfnamefont {D.~G.}\
  \bibnamefont {Greene}}, \bibinfo {author} {\bibfnamefont {D.}~\bibnamefont
  {Hayes}}, \bibinfo {author} {\bibfnamefont {R.}~\bibnamefont {Kroe-Barrett}},
  \bibinfo {author} {\bibfnamefont {T.}~\bibnamefont {Laue}}, \bibinfo {author}
  {\bibfnamefont {J.}~\bibnamefont {Lin}}, \bibinfo {author} {\bibfnamefont
  {B.}~\bibnamefont {McGarry}}, \bibinfo {author} {\bibfnamefont
  {V.}~\bibnamefont {Razinkov}}, \bibinfo {author} {\bibfnamefont
  {S.}~\bibnamefont {Singh}}, \bibinfo {author} {\bibfnamefont
  {R.}~\bibnamefont {Taing}}, \bibinfo {author} {\bibfnamefont
  {S.}~\bibnamefont {Venkataramani}}, \bibinfo {author} {\bibfnamefont
  {W.}~\bibnamefont {Weiss}}, \bibinfo {author} {\bibfnamefont
  {D.}~\bibnamefont {Yang}}, \ and\ \bibinfo {author} {\bibfnamefont {I.~E.}\
  \bibnamefont {Zarraga}},\ }in\ \href {\doibase 10.1021/bk-2015-1201.ch006}
  {\emph {\bibinfo {booktitle} {State-of-the-{Art} and {Emerging}
  {Technologies} for {Therapeutic} {Monoclonal} {Antibody} {Characterization}
  {Volume} 2. {Biopharmaceutical} {Characterization}: {The} {NISTmAb} {Case}
  {Study}}}},\ \bibinfo {series} {{ACS} {Symposium} {Series}}, Vol.\ \bibinfo
  {volume} {1201}\ (\bibinfo  {publisher} {American Chemical Society},\
  \bibinfo {year} {2015})\ pp.\ \bibinfo {pages} {285--327}\BibitemShut
  {NoStop}%
\bibitem [{\citenamefont {Wolter}\ \emph {et~al.}(2015)\citenamefont {Wolter},
  \citenamefont {Pullen}, \citenamefont {Baudisch}, \citenamefont {Sclafani},
  \citenamefont {Hemmer}, \citenamefont {Senftleben}, \citenamefont
  {Schr{\"o}ter}, \citenamefont {Ullrich}, \citenamefont {Moshammer},\ and\
  \citenamefont {Biegert}}]{wolter_strong-field_2015}%
  \BibitemOpen
  \bibfield  {author} {\bibinfo {author} {\bibfnamefont {B.}~\bibnamefont
  {Wolter}}, \bibinfo {author} {\bibfnamefont {M.~G.}\ \bibnamefont {Pullen}},
  \bibinfo {author} {\bibfnamefont {M.}~\bibnamefont {Baudisch}}, \bibinfo
  {author} {\bibfnamefont {M.}~\bibnamefont {Sclafani}}, \bibinfo {author}
  {\bibfnamefont {M.}~\bibnamefont {Hemmer}}, \bibinfo {author} {\bibfnamefont
  {A.}~\bibnamefont {Senftleben}}, \bibinfo {author} {\bibfnamefont {C.~D.}\
  \bibnamefont {Schr{\"o}ter}}, \bibinfo {author} {\bibfnamefont
  {J.}~\bibnamefont {Ullrich}}, \bibinfo {author} {\bibfnamefont
  {R.}~\bibnamefont {Moshammer}}, \ and\ \bibinfo {author} {\bibfnamefont
  {J.}~\bibnamefont {Biegert}},\ }\href {\doibase 10.1103/PhysRevX.5.021034}
  {\bibfield  {journal} {\bibinfo  {journal} {Physical Review X}\ }\textbf
  {\bibinfo {volume} {5}},\ \bibinfo {pages} {021034} (\bibinfo {year}
  {2015})}\BibitemShut {NoStop}%
\end{thebibliography}%
\bibliographystyle{apsrev4-1}

\end{document}